\documentclass[numberedappendix,twocolumn,twocolappendix]{openjournal}

\usepackage{latexsym}
\usepackage{graphicx}
\usepackage{amssymb}
\usepackage{amsmath}
\usepackage{gensymb}
\usepackage[backref=page]{hyperref}
\hypersetup{colorlinks=true,linkcolor=blue,citecolor=blue,filecolor=blue,urlcolor=blue}
\usepackage{cleveref}
\usepackage{bm}		
\usepackage{natbib}

\usepackage[caption=false]{subfig}

\usepackage{tikz}
\usetikzlibrary{positioning}
\usetikzlibrary{decorations.pathreplacing}

\newcommand{\beq}{\begin{eqnarray}}
\newcommand{\eeq}{\end{eqnarray}}
\newcommand{\ben}{\begin{enumerate}}
\newcommand{\een}{\end{enumerate}}
\newcommand{\bit}{\begin{itemize}}
\newcommand{\eit}{\end{itemize}}

\newcommand{\lcdm}{\xspace{\ensuremath{\Lambda\mathrm{CDM}}}\xspace}

\newcommand{\zobs}{z_{\rm obs}}
\newcommand{\tobs}{t_{\rm obs}}

\newcommand{\Msun}{{\rm M}_{\odot}}

\newcommand{\mpeakzero}{m_{\rm p,0}}

\newcommand{\tpeak}{t_{\rm p}}
\newcommand{\Mpeak}{M_{\rm p}}
\newcommand{\Mhalo}{M_{\rm halo}}
\newcommand{\Mh}{M_{\rm h}}
\newcommand{\Mstar}{M_{\star}}
\newcommand{\Mstardot}{\dot{M}_{\star}}

\newcommand{\dstar}{{Diffstar}\xspace}
\newcommand{\Dstar}{{\tt Diffstar}\xspace}

\newcommand{\DstarPop}{{DiffstarPop}\xspace}
\newcommand{\dmah}{{Diffmah}\xspace}

\newcommand{\Fbound}{\mathcal{F}_{\rm bound}}

\newcommand{\thetamah}{\theta_{\rm MAH}}
\newcommand{\uthetamah}{\tilde{\theta}_{\rm MAH}}
\newcommand{\thetasfh}{\theta_{\rm SFH}}
\newcommand{\uthetasfh}{\tilde{\theta}_{\rm SFH}}
\newcommand{\psisfh}{\psi_{\rm SFH}}
\newcommand{\psisfhq}{\psi_{\rm SFH}^{\rm q}}
\newcommand{\psisfhms}{\psi_{\rm SFH}^{\rm ms}}
\newcommand{\psisfhfquench}{\psi_{\rm SFH}^{\rm fqch}}

\newcommand{\Mzero}{M_{0}}

\newcommand{\aearly}{\alpha_{\rm early}}
\newcommand{\alate}{\alpha_{\rm late}}
\newcommand{\tauc}{\tau_{\rm c}}

\newcommand{\Mstart}{M_{\star}(t)}

\newcommand{\mseff}{\epsilon_{\rm ms}}

\newcommand{\sfrtfrac}{\frac{{\rm d}\Mstart}{{\rm d}t}}
\newcommand{\sfrtfracms}{\frac{{\rm d}M_{\star}^{\rm ms}(t)}{{\rm d}t}}

\newcommand{\Mcrit}{M_{\rm crit}}

\newcommand{\Qfunc}{F_{\rm q}}

\newcommand{\qdrop}{q_{\rm drop}}
\newcommand{\qrejuv}{q_{\rm rejuv}}
\newcommand{\qtime}{t_{\rm q}}
\newcommand{\qtimeone}{t_{\rm q_1}}
\newcommand{\qtimetwo}{t_{\rm q_2}}

\newcommand{\qdt}{q_{\rm dt}}

\newcommand{\fquench}{f_{\rm qch}}

\usepackage{xspace}
\newcommand{\tng}{IllustrisTNG\xspace}
\newcommand{\um}{UniverseMachine\xspace}
\newcommand{\galcus}{Galacticus\xspace}

\newcommand{\sclip}{\mathrm{sclip}}

\begin{document}

\title{DiffstarPop: A Generative Physical Model of Galaxy Star Formation History}
\shorttitle{DiffstarPop}

\author{Alex Alarcon$^{1,\star}$, Andrew P. Hearin$^{2}$, Matthew R. Becker$^{2}$,  Gillian Beltz-Mohrmann$^{2,4,5}$, Andrew Benson$^{3}$, Sachi Weerasooriya$^{3}$}

\affiliation{$^1$Institute of Space Sciences (ICE, CSIC), Campus UAB, Carrer de Can Magrans, s/n, 08193 Barcelona, Spain}
\affiliation{$^2$HEP Division, Argonne National Laboratory, 9700 South Cass Avenue, Lemont, IL 60439, USA}
\affiliation{$^3$Carnegie Observatories, 813 Santa Barbara Street, Pasadena, CA 91101, USA}
\affiliation{$^4$Department of Physics, Smith College, 1 Chapin Way, Northampton, MA 01063, USA}
\affiliation{$^5$Program of Statistical and Data Sciences, Smith College, 1 Chapin Way, Northampton, MA 01063, USA}
\thanks{$^{\star}$E-mail:alexalarcongonzalez@gmail.com}

\shortauthors{Alarcon et al}


\begin{abstract}
We present \DstarPop, a differentiable forward model of cosmological populations of galaxy star formation histories (SFH). In the model, individual galaxy SFH is parametrized by Diffstar, which has parameters $\thetasfh$ that have a direct interpretation in terms of galaxy formation physics, such as star formation efficiency and quenching. \DstarPop is a model for the statistical connection between $\thetasfh$ and the mass assembly history (MAH) of dark matter halos.  We have formulated \DstarPop to have the minimal flexibility needed to accurately reproduce the statistical distributions of galaxy SFH predicted by a diverse range of simulations, including the \tng hydrodynamical simulation, the \galcus semi-analytic model, and the \um semi-empirical model. 
Our publicly available code written in JAX includes Monte Carlo generators that supply statistical samples of galaxy assembly histories that mimic the populations seen in each simulation, and can generate SFHs for $10^6$ galaxies in 1.1 CPU-seconds, or 0.03 GPU-seconds.
We conclude the paper with a discussion of applications of \DstarPop, which we are using to generate catalogs of synthetic galaxies populating the merger trees in cosmological N-body simulations.
\end{abstract}

\maketitle

\twocolumngrid

\section{Introduction}
\label{section:intro}

Theoretical models that connect the properties of galaxies with the dark matter halos they inhabit provide a key ingredient in many approaches to making predictions for the galaxy populations observed by cosmological surveys. Predicting the statistical properties of large galaxy populations can be computationally demanding, and models of the galaxy--halo connection are all confronted with a tradeoff between the level of physical detail of the model, and the computational expense of the predictions. As discussed in \citet{wechsler_tinker_2018_galhalo_review}, this tradeoff provides a natural way to organize the landscape of models of the galaxy--halo connection. At one end of this spectrum sit cosmological hydrodynamical simulations \citep[e.g.,][]{Pillepich2018a,schaye_etal15_eagle,dave_etal19}, which achieve the most physically rich predictions for the galaxy population, but only at the enormous computational expense of directly simulating the evolution of baryonic matter and its feedback processes \citep[see][for a recent review]{crain_vandevoort_2023_hydro_review}. At the opposite end of this spectrum are empirical models such as the Halo Occupation Distribution \citep[HOD,][]{Berlind2002,zheng_etal05_hod}, the Conditional Luminosity Function \citep[CLF,][]{yang_mo_vdb_2003_clf,van_den_bosch_etal03_clf}, and subhalo abundance matching \citep[SHAM,][]{kravtsov_etal04,conroy_etal06}; such models are computationally cheap and are based on much less expensive gravity-only N-body simulations, but these models rely on numerous simplifying assumptions that limit their ability to make predictions with physically realistic levels of complexity.

Semi-analytic models (SAMs) strike a balance between these two extremes: their predictions for the galaxy population are physically rich and detailed, but SAMs are much less expensive than hydro simulations. SAMs are based on N-body simulations, or merger trees constructed using extended Press-Schechter theory \citep{lacey_cole_93, jiang_vdb_2014b}, and so do not explicitly solve for the evolution of baryonic mass with tracer particles or grid cells; instead, SAMs employ a set of assumptions about the coupled system of ODEs that regulates the physics of the galaxy that evolves in the merger tree of a simulated halo \citep[for review articles, see][]{benson_2010_galaxy_formation_review,somerville_dave_sam_hydro_review}. For each simulated merger tree, a modern SAM such as Galform \citep{cole_lacey_galform_2000}, Galacticus \citep{benson_galacticus_2012}, GAEA \citep{fontanot_etal24_gaea}, or Shark \citep{lagos_shark_2018} solves an ODE system for the evolution of a reservoir of hot gas that cools and forms stellar mass, for how feedback from star formation ejects cold gas from the galaxy, or for how mergers transform disks into spheroids. SAMs have rich predictive power for multi-wavelength galaxy SEDs, and while considerable progress has been made in recent years in improving their efficiency \citep[e.g.,][]{van_daalen_etal16,henriques_etal_2017}, SAMs are still quite computationally demanding compared to empirical models, and exploring the full parameter space with Bayesian MCMCs can be costly \citep[see, e.g.,][]{henriques_etal09}.

Empirical models of galaxy star formation history (SFH) have become more common in recent years. In this approach, empirical prescriptions are adopted for the relationship between the SFH of a galaxy and the mass assembly history (MAH) of the halo, and observational data is used to constrain the parameters; examples include SMAD \citep{becker_smad_2015}, EMERGE \citep{moster_emerge1}, and \um \citep{Behroozi2019}. This approach is considerably more efficient than most SAMs and permits expansive MCMCs that span the full parameter space, although these models are still more expensive than the HOD/CLF, and require high-resolution N-body simulations with substructure merger trees.

Differentiable programming is rapidly reshaping forward modeling in cosmology and galaxy formation. Early differentiable simulations such as BORG, ELUCID, and BORG-PM \citep{JascheWandelt13BORG, Wang2014ELUCID, Seljak2017, JascheLavaux19BORG} relied on hand-derived analytical derivatives, enabling field-level weak-lensing predictions \citep[e.g. BORG-WL;][]{Porqueres_BORGWL}. Later frameworks such as the FastPM \citep{Feng2016} implementation with Tensorflow, FlowPM \citep{modi_etal21_flowpm}, adopted automatic differentiation (AD), providing machine-accurate, end-to-end gradients without manual derivative derivations. JAX \citep{jax2018github, Frostig2018JAX} further advanced this paradigm through composable, function-transforming AD that scales efficiently to large simulations. Building on JAX, libraries such as \texttt{JAX-COSMO} \citep{jax-cosmo} and \texttt{LINX} \citep{LINX} provide differentiable predictions of core cosmological quantities, from angular power spectra ($C_\ell$) to Big Bang Nucleosynthesis primordial abundances, using analytic approximations or emulation for the CMB power spectrum. Differentiable particle-mesh (PM) solvers such as \texttt{pmwd} \citep{Li_etal22_pmwd, li_etal24_diffsims_adjoint} demonstrate scalable gradient-based inference directly through $N$-body evolution using adjoint methods. Complementing these efforts, \texttt{DISCO-DJ} \citep{hahn_etal24_discodj1} offers a differentiable Einstein–Boltzmann solver, which combined with a PM $N$-body model \citep{List_etal25_discodj2}, enables fast large-scale structure inference, including halo mass functions \citep{buisman_etal25_discodj_hmf}. Fully differentiable hydrodynamics \citep[\texttt{diffHydro};][]{horowitz_etal25_diffhydro} and differentiable halo finders \citep[\texttt{jFoF};][]{horowitz_etal25_diff_fof} extend this approach to gas dynamics and discrete object identification in PM simulations. Finally, \texttt{halox} \citep{halox} provides differentiable dark matter halo properties and large-scale structure calculations for halo modeling.

While PM methods sacrifice small-scale information and therefore underresolve low-mass halos, their performance can be enhanced by incorporating differentiable HOD models \citep[\texttt{DiffHOD};][]{horowitz_etal24_diffhod}.
Alternatively, machine-learning methods can be used to learn mappings from PM simulations to high-resolution $N$-body results, enabling the painting of halos or galaxies \citep{pandey_etal24_charm, pandey_etal24_halo_llm, Pandey15_llm_galactification}.
Another strategy is to use merger trees from a limited set of high-resolution $N$-body simulations, either by differentiating through a semi-analytic model, like in \texttt{sapphire} \citep{Pandya206sapphire}, or by replacing each halo’s history with a differentiable surrogate model such as \dmah \citep{hearin_etal21_diffmah} or Diffprof \citep{stevanovich_etal23_diffprof}. These predictions can then be extended to galaxy SEDs with \dstar and DSPS \citep{alarcon_etal23_diffstar, hearin_etal23_dsps}, and to halo gas content with models such as GODMAX \citep{Pandey_etal25_godmax} or picasso \citep{florian_picasso24}.
Together with cosmological evidence modeling techniques \citep{lange_etal19, lange_etal22} and differentiable large-scale structure summary statistics \citep{hearin_etal21_shamnet}, these developments open the door to deriving cosmological constraints from joint, multi-wavelength observations of multiple tracer populations across cosmic time.

In this paper, we introduce \DstarPop, a new model for the statistical connection between halo MAH and galaxy SFH. \DstarPop relies on a fully parametric model for dark matter halo growth, Diffmah \citep{hearin_etal21_diffmah}, which approximates the MAH of an individual halo with parameters $\thetamah;$ \DstarPop further relies on a fully parametric model for individual galaxy SFH, Diffstar \citep{alarcon_etal23_diffstar}, which is built on top of Diffmah and approximates each galaxy SFH with additional parameters $\thetasfh;$ \DstarPop is a parametric model for $P(\thetasfh\vert\thetamah),$ the probability that a halo with $\thetamah$ hosts a galaxy with $\thetasfh.$ The individual galaxy SFH model, Diffstar, is analogous to other parametric forms used in the literature such as a delayed exponential \citep{Sandage1986_sfhs}, a lognormal \citep{Gladders2013, diemer_etal17}, or a double power-law \citep{Behroozi2013, Ciesla2017, Carnall2018}. However, one feature that distinguishes Diffstar from these and other such models is that each of its parameters $\thetasfh$ have a direct interpretation in terms of galaxy formation physics, such as star formation efficiency, $\mseff,$ and a possible quenching event at time $\qtime.$ Another important difference is that \dstar is built upon \dmah, linking galaxy growth directly to the underlying halo mass assembly history. 

\DstarPop leverages the Diffstar formulation of the SFH--MAH relationship to provide a physics-based parameterization of galaxy--halo connection. A key technical advance of our approach is that \DstarPop is a fully differentiable model implemented in the JAX autodiff library, which enables us to use gradient-based algorithms for optimization and inference, and to harness the computational power of GPUs. As we show below, this aspect of our model plays a central role in the principal aim of this paper: stress-testing the flexibility of \DstarPop to reproduce the statistical population of galaxy SFHs predicted by three of the leading models in the field, \tng, \galcus, and \um. 

This paper is organized as follows. In \S\ref{section:sims} we describe the simulations we use in this work. In \S\ref{section:model_form} we describe the DiffstarPop galaxy-halo connection model in detail. In \S\ref{section:results} we show the results. We discuss our findings in \S\ref{section:discussion}, and we conclude in \S\ref{section:conclusions} with a summary of our primary results. The appendices contain more granular technical information than the main body, as well as a more comprehensive collection of validation plots.

\section{Simulated Galaxies}
\label{section:sims}
In this paper, we stress-test the flexibility of \DstarPop by assessing its ability to approximate statistical distributions of galaxy SFHs seen in three very different simulated populations: the \tng hydrodynamical simulation \citep{Pillepich2018a}, the \galcus semi-analytic model \citep{benson_galacticus_2012}, and the \um semi-empirical model \citep{behroozi_etal19}. In this section, we describe each of these sources of simulated SFHs in turn.

\tng is a suite of hydrodynamical simulations that models how dark matter, gas, and stars evolve in a cosmological context \citep{Weinberger2017, Naiman2018, Springel2018, Marinacci2018, Pillepich2018a, Pillepich2018b, Pillepich2019, Nelson2018, Nelson2019}. We use publicly available data from TNG300-1 \citep{Nelson2018a}, which was generated by using {\sc Arepo} \citep{Springel2010} to solve for the evolution of $2500^3$ gas tracers together with the same number of dark matter particles in a periodic box of $302.6$ Mpc on a side evolving under {\it Planck} 2015 cosmology. The mass resolution for dark matter and gas is 5.9 and $1.1\times10^7\Msun$, respectively. For the \tng halo catalogs and merger trees we use in this work,
halos and subhalos were identified with {\tt SUBFIND} \citep{springel_etal2001_subfind}, and merger treers were constructed with {\tt SUBLINK} \citep{rodriguez_gomez_etal15_sublink}. The cosmology adopted follows \textit{Planck} 2015 results \citep{Planck15}, with $H_0=67.74\,\mathrm{km\,s^{-1}\,Mpc^{-1}}$, $\Omega_\mathrm{m}=0.3089$, $\Omega_\Lambda=0.6911$, and $\Omega_\mathrm{b}=0.0486$. The \tng model was not calibrated through a formal fit to a single observational likelihood, but was instead tuned to reproduce a set of benchmark observables that are predominantly at low redshift ($z=0 - 0.2$), including the galaxy stellar mass function, stellar-to-halo mass relation, galaxy size--mass relation, halo gas fractions, and black hole scaling relations, while also being checked against the redshift evolution of the cosmic star formation rate density as a function of time, $0 < z < 8$ \citep{Weinberger2017, Pillepich2018a}, and the $z<0.1$ SDSS $g-r$ color bimodality \citep{Nelson2018}.

\galcus is a semi-analytic model of galaxy formation that predicts the star formation history of a galaxy by numerically solving a coupled ODE system that regulates how gas and stars evolve in the gravitational potential of a dark matter halo \citep{benson_galacticus_2012}. To generate the halo merger trees for \galcus, we use the Extended Press Schechter (EPS) formalism \citep{press_schechter_1974,bower_1991,bond_cole_efsthathiou_1991,parkinson_cole_helly_2008}, in particular the EPS trees developed in \cite{benson_ludlow_cole_2019_eps_concentration} and implemented in \galcus. 
The merger trees are constructed for halos with root masses in the range $1.0\times10^{11}\Msun$ to $1.78\times10^{14}\Msun$, with a minimum progenitor mass resolution of $2.5\times10^{9}\Msun$ and a fractional mass resolution of $8.33\times10^{-5}$ relative to the root halo mass. The cosmology adopted follows \textit{Planck} 2018 results \citep{Planck18}, with $H_0=67.36\,\mathrm{km\,s^{-1}\,Mpc^{-1}}$, $\Omega_\mathrm{m}=0.3153$, $\Omega_\Lambda=0.6847$, and $\Omega_\mathrm{b}=0.0493$. The \galcus model parameters were chosen by manually searching parameter space for reasonable agreement with a variety of predominantly low-redshift observations, including the $z=0$ stellar mass function, $z=0$ $K$- and $b_J$-band luminosity functions, the local Tully--Fisher relation, the local colour--magnitude distribution, the $z=0$ disc size distribution, the black hole--bulge mass relation, and the cosmic star formation history \citep{benson_galacticus_2012, Knebe2018}.

For \um, we use a population of simulated halos from the Small MultiDark Planck (SMDPL) simulation \citep{klypin_etal16}.
The SMDPL simulation was run using the \textsc{l-gadget-2} code, a modified version of the publicly available \textsc{gadget-2} code \citep{springel_2005_gadget2}. The simulation was run by gravitationally evolving dark matter particles of mass $8.14\times10^8\Msun$ in a periodic box with side length of 590 Mpc in a flat \lcdm cosmology with the following parameters: $\Omega_\mathrm{m}=0.307$, $\Omega_{\Lambda}=0.693$, $\Omega_\mathrm{b}=0.048$, $h=0.6777$, $\sigma_8=0.8228$, and $n_s=0.96$ \citep{planck14b}.
Dark matter (sub)halos were identified using \textsc{rockstar} \citep{behroozi_etal13_rockstar}, and merger trees were generated using \textsc{Consistent Trees} \citep{behroozi_etal13_consistent_trees}. Galaxy SFHs were generated by running the publicly available \um source code\footnote{\url{https://bitbucket.org/pbehroozi/universemachine}} with the default best-fitting parameters from DR1 associated with \citet{behroozi_etal19}. The \um model was calibrated directly to a broad set of observational constraints spanning multiple redshifts, including stellar mass functions ($0\lesssim z\lesssim 4$), cosmic star formation rates ($0\lesssim z\lesssim 10$), specific star formation rates ($0\lesssim z\lesssim 8$), UV luminosity functions ($4\lesssim z\lesssim 10$), and quenched fractions ($0\lesssim z\lesssim 4$), together with clustering, weak lensing, and environmental quenching constraints where available \citep{behroozi_etal19}.

For the case of the \um and \galcus models, the model formulation makes it simple to separately predict and extract the {\em in-situ} component of star formation, which includes only stellar mass formed by integrating the main progenitor of the (sub)halo, from the {\em ex-situ} component, which includes a contribution from stellar mass brought in with merged satellite galaxies. For \tng, the in-situ and ex-situ components of star formation are not as cleanly separable or as straightforward to extract. We adopt the same definitions of in-situ and ex-situ stars as used in \citet{Rodriguez-Gomez16, Pillepich2018b,Tacchella2019}, based on the baryonic merger tree reconstruction method of {\tt SUBLINK}, and in our results for \tng we only consider in-situ SFH. Separating in-situ and ex-situ contributions to total stellar mass assembly will be essential for addressing major mergers in future work (see \S5 for details).

Throughout this paper, we quote numerical values for halo mass $\Mhalo$ and stellar mass $\Mstar$ assuming the value of $h$ in the simulation under consideration, using units of $\Msun$ everywhere. For example, in a simulation with $h=0.6777$ that listed $\Mstar=1\times10^{10}\Msun/h$, we would quote $\Mstar\simeq1.48\times10^{10}\Msun$. The \dmah fits in our analysis were each derived based on $\Mhalo(t)$ assuming the value of $h$ and $t_0$ in the cosmology of the simulation, where $t_0$ is the redshift-zero age of the universe; the \dstar fits to $\Mstardot(t)$ similarly assume $h$ in the simulated cosmology, as well as the cosmic baryon fraction, $f_{\rm b}.$

In all equations and figures, we use the notation log(x) to denote the base-10 logarithm.

\section{DiffstarPop Model Formulation}
\label{section:model_form}

In this section, we give a high-level overview of the DiffstarPop formulation of the galaxy--halo connection; the appendices contain a complete technical description of the model, as well as the validation exercises we have carried out to stress-test our formulation.

In \S\ref{subsec:diffmah_model}, we outline the \dmah model for the mass assembly (MAH) of an individual dark matter halo, $\Mhalo(t);$ Diffmah is a fully parametric model with behavior determined by parameters $\thetamah.$
In \S\ref{subsec:diffstar_model}, we describe the Diffstar model for $\Mstardot(t),$ the star formation history (SFH) of an individual galaxy. 
\dstar is a fully parametric SFH model with parameters $\thetasfh$ that have a direct interpretation in terms of galaxy formation physics.
Both \dmah and \dstar are models for the \emph{individual} growth histories of halos and galaxy SFHs, respectively, and previous work has shown that they are flexible enough to describe individual galaxies and halos across a range of simulations \citep{hearin_etal21_diffmah, alarcon_etal23_diffstar}.
\DstarPop is a model for $P(\thetasfh\vert\thetamah),$ the \emph{population} of Diffstar galaxies that occupy dark matter halos. \DstarPop has its own parameters, $\psisfh,$ that determine how the PDF of $\thetasfh$ varies with $\thetamah.$ 

In Section~\ref{subsec:diffstarpop_model}, we outline the \DstarPop parametrization of $P_{\psisfh}(\thetasfh\vert\thetamah)$. The functional forms defining \DstarPop are motivated by relations between \dstar parameters and halo properties, specifically halo mass and growth history. Our goal is to construct a minimally flexible model that robustly captures key summary statistics of the galaxy--halo PDF: the stellar-to-halo-mass PDF as a function of redshift, and the PDF of specific star formation rates, as a function of stellar mass and redshift. Results and definitions for these PDFs are shown in \S\ref{section:results}.

\subsection{Diffmah model of individual halo MAH}
\label{subsec:diffmah_model}
\dmah is a parametric model for the mass assembly history (MAH) of individual halos. In particular, \dmah approximates the evolution of {\em cumulative peak} halo mass, $\Mpeak(t),$ defined as the largest mass the main progenitor halo has ever attained up until the time $t,$ so that $\Mpeak(t)$ is a non-decreasing function for all $t.$ For a typical central halo, $\Mpeak(t)\approx M_{\rm halo}(t),$ but if a halo experiences either a temporary or sustained period of mass loss, $\Mpeak(t)$ remains constant.

We model $\Mpeak(t)$ with a power-law function of cosmic time with a rolling index,
\beq
\label{eq:individual_mah}
\Mpeak(t) = \Mzero(t/t_0)^{\alpha(t)},
\eeq
where $t_0$ is the present-day age of the universe, $\Mzero$ is a normalization parameter, and the behavior of $\alpha(t)$ is controlled by a sigmoid function:
\beq
\label{eq:sigmoid}
\alpha(t) = \aearly + \frac{\alate-\aearly}{1+\exp(-k(t-\tauc))}.
\eeq
At early times, $t\ll\tauc,$ the power law slope $\alpha(t)\approx\aearly,$ and conversely, $\alpha(t)\approx\alate$ at late times $t\gg\tauc;$ the sigmoid function in Eq.~\ref{eq:sigmoid} smoothly transitions $\alpha(t)$ between these two regimes.

\begin{figure}
\includegraphics[width=\columnwidth]{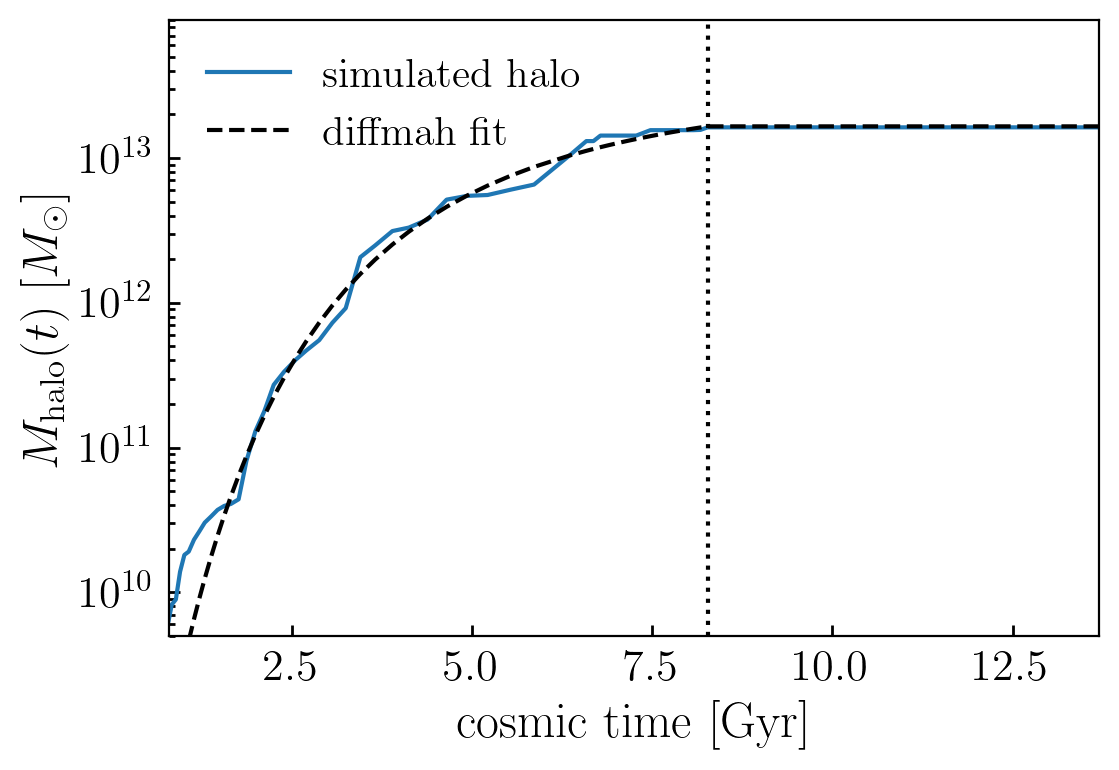}
\caption{{\bf Diffmah model of individual halo growth.} The solid blue curve shows the mass assembly history (MAH) of an individual dark matter halo in the SMDPL simulation. The dashed black curve shows the Diffmah approximation of the simulated halo MAH. The dotted vertical line shows $\tpeak,$ the Diffmah parameter encoding the time at which halo growth is arrested.}
\label{fig:diffmah_example_fit}
\end{figure}

In \citet{hearin_etal21_diffmah}, the above model was fit to millions of simulated merger trees in both gravitational N-body simulations as well as \tng, and it was shown that the Diffmah model gives an unbiased fit with $0.1$dex of scatter across a wide range of halo mass, $10^{11}\lesssim \Mhalo/\Msun\lesssim10^{15},$ and redshift, $0\leq z\lesssim5.$ In this paper, we introduce a new Diffmah parameter, $\tpeak,$ defined as the cosmic time when the mass of the halo peaks, and remains constant thereafter:
\beq
\label{eq_tpeak_def}
\Mpeak(t) = \begin{cases}
    \Mzero(t/t_0)^{\alpha(t)} & t\leq \tpeak \\
    \Mzero(\tpeak/t_0)^{\alpha(\tpeak)} & t\geq\tpeak
\end{cases}
\eeq
As described in Appendix~\ref{diffmah_appendix}, the new parameter $\tpeak$ improves the quality of the Diffmah fits for subhalos with early infall times. We define $\tpeak$ as the first snapshot at which the maximum $\Mpeak$ is attained for a given halo (see Appendix~\ref{diffmah_appendix} for more details).  Figure~\ref{fig:diffmah_example_fit} shows a visual demonstration of an example Diffmah fit to a particular halo in the SMDPL simulation described in \S\ref{section:sims}.

We refer the reader to \citet{hearin_etal21_diffmah} and Appendix~\ref{diffmah_appendix} for further details about the Diffmah model.

\subsection{Diffstar model of individual galaxy SFH}
\label{subsec:diffstar_model}

\dstar is a fully parametric model for $\Mstardot(t),$ the star formation history (SFH) of an individual galaxy, introduced by \citet{alarcon_etal23_diffstar}. In this work, we present an updated version of the model, that substantially improves its accuracy, speed (by a factor of 10), and memory efficiency (by a factor of 10). Here, we summarize the physical basis and structure of the \dstar model, while in Appendix~\ref{diffstar_appendix} we give more details of the changes introduced, as well as demonstrations of its flexibility and accuracy.

In the basic physical picture underlying \dstar, 
baryonic matter becomes available for star formation at a rate determined by the growth rate of the dark matter halo:
\beq
\label{eq:dmgasdt}
M_{\rm g}^{\rm acc}(t) = f_{\rm b}\, \Mpeak(t),
\eeq
where $f_{\rm b}$ is the cosmic baryon fraction, $M_{\rm g}^{\rm acc}(t)$ is the cumulative accretion history of baryonic material from the field, and $\Mpeak(t)$ is the halo mass assembly history. In \dstar, we use the \dmah model to approximate $\Mpeak(t)$, Eq.~\ref{eq:individual_mah}.

Once gas is accreted into the dark matter halo, we assume that only a fraction of it is converted into stars, and that this conversion is a gradual process that may take several Gyr \cite[e.g.][]{semenov_etal17_sfr_efficiency1, semenov_etal17_sfr_efficiency2}. 
In formulating $\Mstardot(t)$, we follow earlier empirical models in which the star formation rate is written as a halo-growth term, specifying when new baryonic material becomes available for star formation, multiplied by a baryon-conversion efficiency that controls how effectively this material is converted into stars \citep{BehrooziWechsler2013,Mutch2013,moster_emerge1, alarcon_etal23_diffstar}. Here we approximate the star formation rate of a main sequence galaxy by multiplying the cumulative accretion history of baryonic material from the field, $M_{\rm g}^{\rm acc}(t)$, by an efficiency parameter with units of inverse time,  $\mseff$:  
\beq
\label{eq:mssfr}
\sfrtfracms=\mseff(\Mpeak(t)) \, M_{\rm g}^{\rm acc}(t)
\eeq
For instance, a galaxy that at a given time $t$ has accreted a total of $M_{\rm g}^{\rm acc}(t)=10^{10} \,\Msun$ and has an efficiency of $\mseff= 10^{-9} \,\rm{yr}^{-1}$, would have a star formation rate of ${\rm d}M_{\star}^{\rm ms}(t)/{\rm d}t = 10 \,\Msun/\rm{yr}$.

Motivated by \citet{Mutch2013}, the original \dstar model, and the finding of \citet{BehrooziWechsler2013} that star formation efficiency depends much more strongly on halo mass than on time, we model $\mseff$ as a function of $\Mpeak(t)$ alone. We assume that $\mseff(\Mpeak)$ has a characteristic peak and declines toward both lower and higher halo masses, and we adopt an analytic form identical to the original \dstar model, and similar to that used in \citet{moster_emerge1}:
\beq
\label{eq:mseff}
\mseff(\Mpeak) = \epsilon_{\rm crit}\cdot(\Mpeak/\Mcrit)^{\beta(\Mpeak)}
\eeq
In Eq.~\ref{eq:mseff}, we model $\beta(\Mpeak)$ with another sigmoid (see Eq.~\ref{eq:sigmoid}). The efficiency attains its critical value of $\epsilon_{\rm crit}$ when $\Mpeak=\Mcrit,$ where $\Mcrit\approx10^{12}M_{\odot}$ for typical galaxies.
Although we do not introduce explicit parameter dependence on redshift or halo assembly history in the baryon-conversion efficiency, as in EMERGE \citep{moster_emerge1}, or in the SFR itself, as in UniverseMachine \citep{Behroozi2019}, assembly information still enters our model through the full accretion history encoded in $M_{\rm g}^{\rm acc}(t)$. We find that this leading-order approximation is sufficient to describe the SFHs in our target simulations.

Some galaxies experience a quenching event that shuts down star formation; in \dstar, we capture this phenomenon through the {\it quenching function,} $\Qfunc(t),$ which acts as a multiplicative factor on the star formation rate:
\beq
\label{eq:quenching}
\sfrtfrac = \Qfunc(t)\times\sfrtfracms.
\eeq

\begin{figure}
\includegraphics[width=\columnwidth]{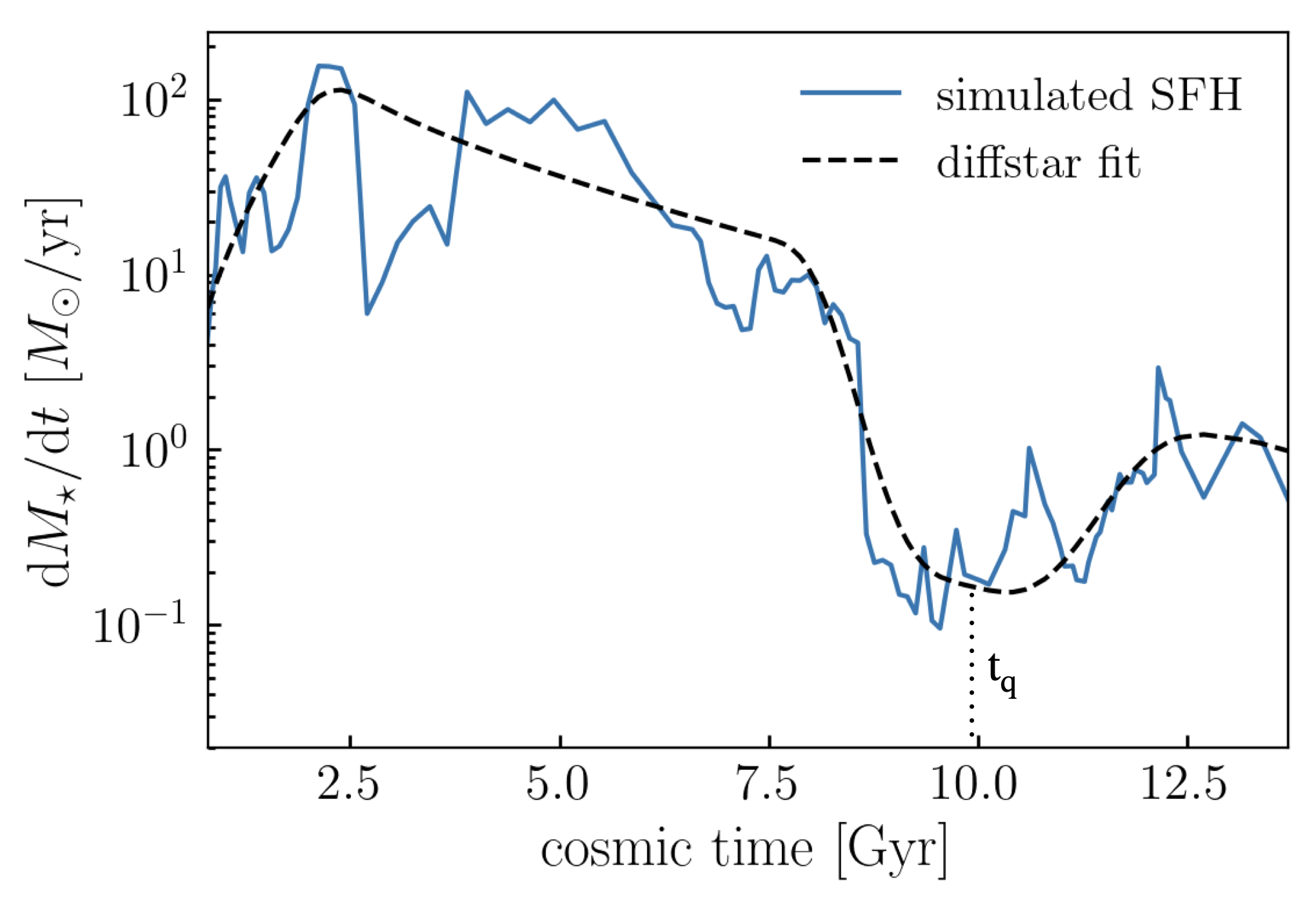}
\caption{{\bf Diffstar model of individual galaxy SFH.} The solid blue curve shows the SFH of an individual UniverseMachine galaxy. The dashed black curve shows the Diffstar approximation. The vertical dotted line indicates the quenching time.}
\label{fig:diffstar_example_fit}
\end{figure}

As described in Section 3.5 in \citet{alarcon_etal23_diffstar}, we use a sigmoid function to define the behavior of $\Qfunc(t)$ as a smooth logarithmic drop in star formation rate. To parameterize a quenching event in \dstar, the parameter $\qtime$ is the time at which the quenching event reaches $\qdrop,$ its maximum suppression of SFR; an additional parameter $\qdt$ controls how quickly the SFR drops to its asymptotic value. The rejuvenation of star formation of previously-quenched galaxies has been shown to be a rather common phenomenon in the Magneticum hydrodynamical simulation \citep{fortune_etal25_magneticum_sfh}, as well as in UniverseMachine \citep{alarcon_etal23_diffstar}.
Observations indicate that rejuvenation events contribute a small fraction ($<10\%$) of the total stellar mass of a galaxy, that they might be more likely in disk-like galaxies,  and the rejuvenation fraction of different samples ranges from 10\% to 30\% depending on the selection method \citep{Chauke2019,Mancini2019,Tacchella2022,Zhang2023,Tanaka19_HINOTORI,Wan2025}.
To capture this effect in \dstar, a quenched galaxy has parametric freedom to experience rejuvenation that drives its SFR back to the main sequence: the parameter $\qrejuv$ controls the level of rejuvenation, while $\qdt$ again controls the timescale over which the SFR rises to its $\qrejuv$ value.

Figure~\ref{fig:diffstar_example_fit} demonstrates an example fit of Diffstar to a particular galaxy in UniverseMachine. In \citet{alarcon_etal23_diffstar}, it was shown that \dstar is sufficiently flexible to accurately approximate simulated SFHs in \um and \tng. We give further demonstrations of this flexibility in Appendix~\ref{diffstar_appendix}, in which we show that \dstar is also able to give a good approximation to SFHs in \galcus.

\subsection{\DstarPop model of the SFH of galaxy populations}
\label{subsec:diffstarpop_model}

\begin{figure}
\centering
\begin{tikzpicture}[
  node distance=0.75cm,
  box/.style={draw, rounded corners, align=center, inner sep=4pt,
              font=\scriptsize, text width=2.6cm},
  box2/.style={draw, rounded corners, align=center, inner sep=4pt,
              font=\scriptsize, text width=2.0cm, fill=gray!12},
  box3/.style={draw, rounded corners, align=center, inner sep=4pt,
              font=\scriptsize, text width=2.0cm},
  pop/.style={draw, rounded corners, align=center, inner sep=4pt,
              font=\scriptsize, text width=3.2cm, fill=gray!12},
  pop2/.style={draw, rounded corners, align=center, inner sep=4pt,
              font=\scriptsize, text width=2.6cm, fill=gray!12},
  arrow/.style={->, thick},
  darrow/.style={->, thick, dashed},
  ddarrow/.style={->, thick, dotted},
]

\node[box] (thetaMAH) {$\theta_{\rm MAH}$\\
Diffmah params\\
incl. $t_p$};

\node[box, below=of thetaMAH] (Mp) {$M_p(t)$\\halo MAH $\rightarrow \mpeakzero$};

\node[box, below=of Mp] (Mg) {$M_g^{\rm acc}(t)=f_bM_p(t)$\\
accreted baryons};

\node[pop, right=of Mp] (gmm)
{$P(\theta_{\rm SFH}\mid\theta_{\rm MAH},\psi_{\rm SFH})$\\
$=(1-f_{\rm qch})P_{\rm ms}+f_{\rm qch}P_q$};


\node[pop2, below right=0.5cm and -2.9cm of gmm] (fqch)
{$f_{\rm qch}(\mpeakzero,\tpeak)$\\
quenched fraction};

\node[pop2, below right=1.5cm and -2.9cm of gmm] (Pms)
{$P_{\rm ms}$\\
main sequence\\
4 efficiency params\\
};

\node[pop2, below right=3.5cm and -2.9cm of gmm] (Pq)
{$P_q$\\
quenched\\
4 efficiency + 4 quenching params\\
};

\node[box2, below right=0.5cm and -2.3cm of Mg] (thetaSFH)
{sample $\theta_{\rm SFH}$};

\node[box3, below=1.5cm of Mg] (eps)
{$\epsilon_{\rm ms}(M_p(t))$\\
main-sequence efficiency};

\node[box3, below=of eps] (sfrms)
{$\dot M_\star^{\rm ms}(t)=\epsilon_{\rm ms} \times M_g^{\rm acc} $};

\node[box3, below right=0.5cm and -1.9cm of sfrms] (Fq)
{$F_q(t)$\\
quenching / rejuvenation};

\node[box3, below=4.0cm of eps] (sfh)
{$\dot M_\star(t)$\\
galaxy SFH};

\draw[arrow] (thetaMAH) -- (Mp);
\draw[arrow] (Mp) -- (Mg);

\draw[darrow] (gmm.south west) -- ++(0.2,0) |- (fqch.west);
\draw[darrow] (gmm.south west) -- ++(0.2,0) |- (Pms.west);
\draw[darrow] (gmm.south west) -- ++(0.2,0) |- (Pq.west);

\draw[arrow] (gmm.south west) --  ++ (-0.2,0) |- (thetaSFH.east);

\draw[arrow] (thetaSFH.south) +(-0.3,0.0) -- (eps.north);
\draw[arrow] (eps) -- (sfrms);
\draw[arrow] (thetaSFH.south east) +(-0.1,0.0) -- ([xshift=-0.2cm]Fq.north east);

\draw[arrow] (sfrms.south west) +(0.2,0.0) -- ([xshift=0.2cm]sfh.north west);
\draw[arrow] (Fq.south) +(-0.4,0.0) -- (sfh.north);

\draw[darrow] (thetaMAH.east) -- ++(+0.5,0) |- (gmm.west);
\draw[darrow] (Mp.east) -- ++(+0.5,0) |- (gmm.west);
\draw[darrow] (Mp.south west) -- ++(-0.1,0) |- (eps.west);
\draw[arrow] (Mg.south west) -- ++(0.0,0) |- (sfrms.west);

\node[font=\scriptsize, align=center, below=0.15cm of Pms]
{means/scatters mainly \\ depend on $m_{p,0}$};

\node[font=\scriptsize, align=center, below=0.15cm of Pq]
{means/scatters\\mainly depend on $m_{p,0}$;\\
$\langle t_q\rangle$ shift
depends on \\ $m_{p,0}$ and $t_p$};

\draw[decorate, decoration={brace, amplitude=3pt, mirror}]
  ([xshift=-0.25cm]thetaMAH.north west) -- ([xshift=-0.25cm]Mg.south west)
  node[midway, left=8pt, rotate=90, anchor=center, font=\scriptsize]
  {Diffmah};
  
\draw[decorate, decoration={brace, amplitude=3pt, mirror}]
  ([xshift=-0.5cm]eps.north west) -- ([xshift=-0.5cm]sfh.south west)
  node[midway, left=8pt, rotate=90, anchor=center, font=\scriptsize]
  {Diffstar};

  \draw[decorate, decoration={brace, amplitude=5pt}]
  (gmm.north west) -- (gmm.north east)
  node[midway, above=6pt, font=\scriptsize]
  {DiffstarPop};

\end{tikzpicture}
\caption{
Schematic dependency flow of DiffstarPop. Diffmah parameters
$\theta_{\rm MAH}$ (incl. $\tpeak$) determine the halo mass assembly history ($M_p(t)$ and $\mpeakzero$), while DiffstarPop samples $\theta_{\rm SFH}$ from a two-component Gaussian mixture. Most scaling relations in DiffstarPop depend on $m_{p,0}$, whereas the quenched fraction $f_{\rm qch}$ and the mean quenching time $\langle t_q\rangle$ additionally depend on $t_p$. For more details see Section~\ref{subsec:diffstarpop_model} and Appendix~\ref{diffstarpop_appendix}.
}
\label{fig:diffstarpop_flowchart}
\end{figure}

\begin{figure}
\includegraphics[width=\columnwidth]{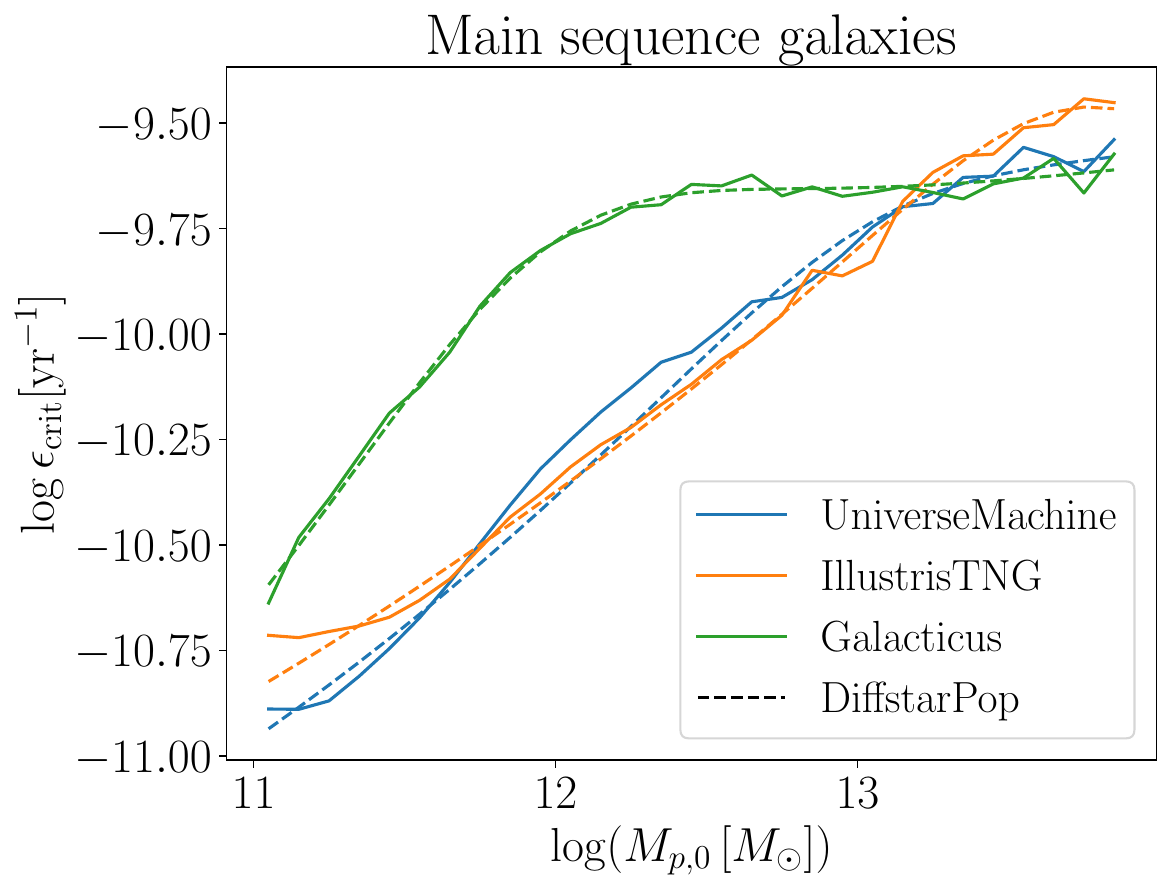}
\caption{{\bf Average critical main sequence efficiency vs. halo mass.} The solid curves show the average value of $\log\epsilon_{\rm crit}$ as a function of present-day halo mass $\mpeakzero$ for \um, \tng and \galcus galaxies, obtained by fitting \dstar to individual galaxies. The dashed curves shows the DiffstarPop approximation of this scaling relation, which are later used as the initial guess for our gradient descent optimizations. DiffstarPop contains ingredients capturing scaling relations such as this one for each of the eight parameters in $\thetasfh$, for both the mean and standard deviation.}
\label{fig:diffstarpop_mean_ecrit}
\end{figure}

\begin{figure}
\includegraphics[width=\columnwidth]{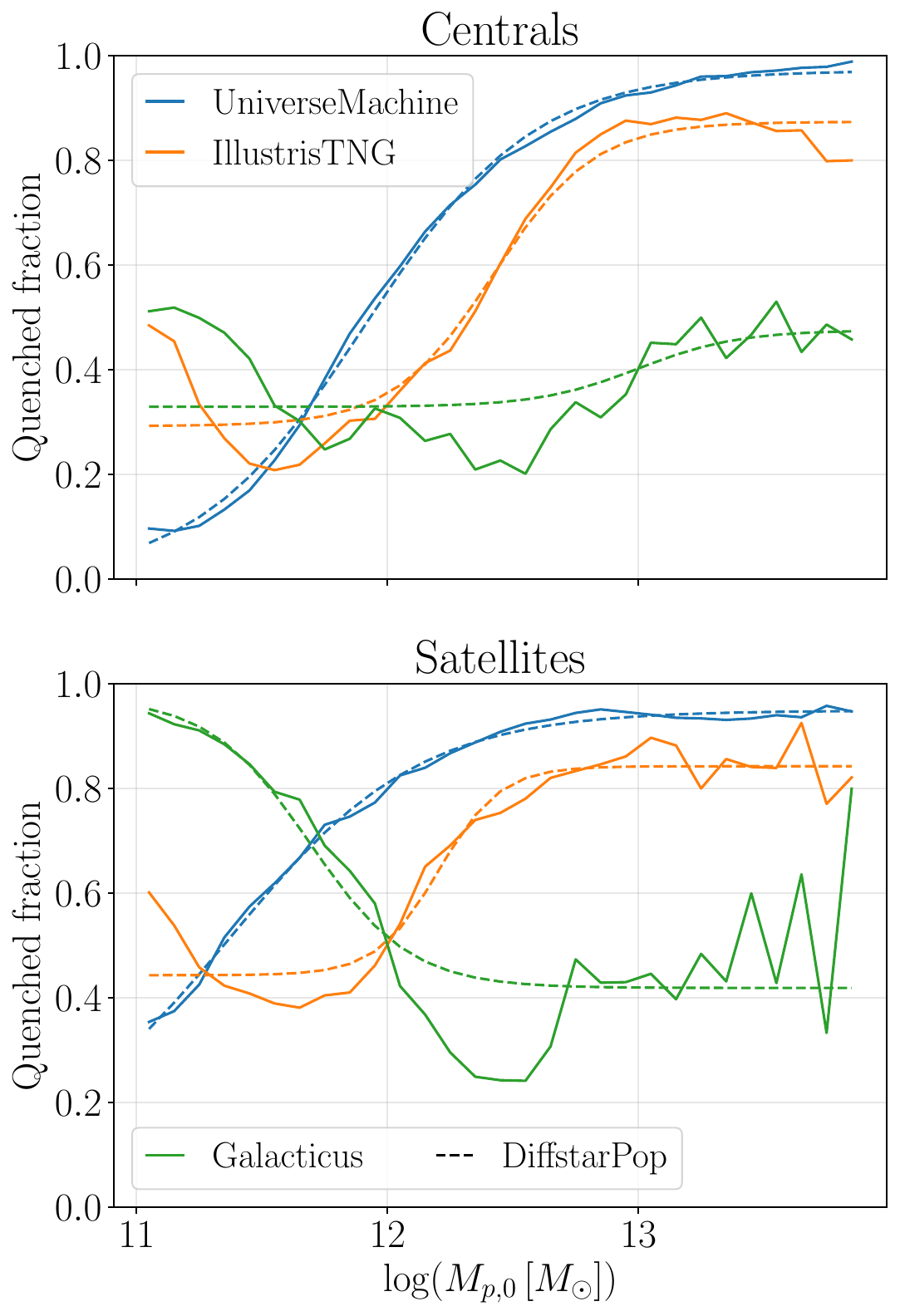}
\caption{{\bf Quenched fraction vs. halo mass.} The solid curves show the average value of the fraction of galaxies that experience a quenching event (i.e. $\qtime<t_0$) as a function of present-day halo mass $\mpeakzero$ for \um, \tng and \galcus galaxies (using only in-situ SFH). The dashed curves show the DiffstarPop approximation of this scaling relation, which are later used as the initial guess for our gradient descent optimizations. Central galaxies are shown on the top panel, and satellites on the bottom panel.}
\label{fig:diffstarpop_mean_qfrac}
\end{figure}

In this section, we give a high-level overview of the formulation of \DstarPop. Figure~\ref{fig:diffstarpop_flowchart} presents a summary flowchart of DiffstarPop and its different components. For a more detailed description of the model, refer to Appendix~\ref{diffstarpop_appendix} and the documentation of our publicly available source code\footnote{\url{https://github.com/ArgonneCPAC/diffstar}}.

\DstarPop is a parametric model for $P(\thetasfh\vert\thetamah),$ the probability that a dark matter halo with MAH given by $\thetamah$ hosts a galaxy with SFH given by $\thetasfh.$ With a model for $P(\thetasfh\vert\thetamah),$ one can directly populate each merger tree of a cosmological N-body simulation with a galaxy SFH, which is one of our primary applications.

We model $P(\thetasfh\vert\thetamah)$ as a two-component multivariate normal distribution, consisting of a ``quenched" population of galaxies that experiences a significant quenching event, and a ``main sequence" population that never experiences such quenching.  The \DstarPop parameters $\psisfh$ encode how the mean and covariance of $P_{\rm ms}(\thetasfh\vert\thetamah)$ and $P_{\rm q}(\thetasfh\vert\thetamah)$ vary with halo properties $\thetamah.$

In \dstar, there are eight parameters $\thetasfh$: four parameters specifying the sigmoid we adopt for $\mseff(\Mpeak)$, and four parameters defining a possible quenching and rejuvenation event $\Qfunc(t).$ For each of these eight parameters, we adopt a scaling relation for how the mean and the standard deviation of the parameter scales with present-day halo mass, $\mpeakzero\equiv\log_{10}(\Mpeak(t_0)).$ 
To model the scaling of the mean relations $\langle\, \cdot \mid \mpeakzero \rangle$, we generally adopt a linear dependence on $\mpeakzero$, smoothly clipped at fixed bounds.  We make three exceptions: the mean relations of $M_{\rm crit}$, $\epsilon_{\rm crit}$, and $\qtime$, for which we instead use a \emph{sigslope} model, in which the slope of the linear relation transitions smoothly between two values through a sigmoid function. Figure~\ref{fig:diffstarpop_mean_ecrit} provides an example, illustrating how $\epsilon_{\rm crit}$ varies with $\mpeakzero$ in the \DstarPop\ model, and also in the best-fitting $\thetasfh$ taken directly from the simulations. Further implementation details are provided in Appendix~\ref{diffstarpop_appendix}.

Since we have four parameters controlling the main sequence efficiency, there are four scaling relations controlling $\langle\mseff(\Mpeak)\vert\mpeakzero\rangle$. However, we have independent scaling relations of $\mseff(\Mpeak)$ for our two subpopulations, ``main sequence" or  ``quenched", which are defined based on whether their \dstar quenching time parameter $\qtime$ is lower/greater than the present time $t_0$. Then, there are four scaling relations controlling $\langle\mseff(\Mpeak)\vert\mpeakzero, \qtime>t_0\rangle$, for the main sequence population, and four scaling relations controlling  $\langle\mseff(\Mpeak)\vert\mpeakzero, \qtime<t_0\rangle$ for the quenched population. For the quenched population, there are an additional four scaling relations that control $\langle\Qfunc(t)\vert\mpeakzero\rangle.$ For the covariance of the quenched and main sequence multivariate normal distributions, we assume there is no correlation between the different parameters, and the scaling relation of the standard deviation of different parameters, e.g. $\sigma(\log\epsilon_{\rm crit}\vert\mpeakzero)$, has a linear dependence with $\mpeakzero$.

As our model for $P(\thetasfh\vert\thetamah)$ is a bimodal distribution, it is also necessary to specify the relative weighting of the two populations, i.e., the {\em quenched fraction}, varies with halo properties. We model the quenched fraction with a sigmoid function of $\mpeakzero$. 
Figure~\ref{fig:diffstarpop_mean_qfrac} shows a comparison between \DstarPop and simulations for the quenched fraction as a function of $\mpeakzero$, for both central and satellite galaxies. The dashed curves show the DiffstarPop approximation of these different scaling relations, which are later used as the initial guess for our gradient descent optimizations run on each simulation. We note that the three simulations exhibit substantial differences in their quenched fractions, particularly for massive galaxies and for satellites. In particular, Galacticus predicts significantly lower quenched fractions for massive centrals and satellites than either \tng\ or UM. We have directly verified that these differences are present in the underlying simulation data, and therefore they reflect genuine differences among the simulated galaxy populations, which our model is flexible enough to capture.
For more technical details of the formulation and implementation, we refer the interested reader to Appendix~\ref{diffstarpop_appendix} and our publicly available code. 

In addition, we introduce an explicit dependence in \DstarPop on a secondary halo property: the Diffmah parameter $\tpeak$. As discussed earlier, this parameter represents the time when the halo mass peaks, and serves as a tracer of halos that experience ``arrested development" \citep[see, e.g.,][]{smith_etal24_arrested_mahs}. We incorporate $\tpeak$-dependence into two separate aspects of the model.
First, we modify the quenched fraction by allowing two of its sigmoid parameters ($y_{\rm lo}$ and $x_0$) to themselves be sigmoids that vary with $\tpeak$. Second, we model the mean of the quenching time $\langle\qtime\vert\mpeakzero\rangle$ with two components: the corresponding linear relationship with $\mpeakzero$, as for all other parameters; and with an additional term that shifts the quenching time, either positively or negatively, based on both $\tpeak$ and $\mpeakzero$ for each halo (see details in Appendix~\ref{sec:subsec_tpeak}).

\section{Results}
\label{section:results}
In \S\ref{subsec:diffstarpop_model}, we gave an overview of how the \DstarPop parameters $\psisfh$ can be used to make predictions for the galaxy--halo PDF, $P(\thetasfh\vert\thetamah).$ In this section, we give some example demonstrations of the results of fitting \DstarPop to the target PDFs:
\ben
\label{eq:targetpdfs}
\item $P(\Mstar\vert\Mh, \zobs)$, the stellar-to-halo-mass PDF as a function of redshift, for all galaxies;
\item $P({\rm sSFR}\vert\Mstar, \zobs)$, the PDF of specific star formation rates, ${\rm sSFR}\equiv\log_{10}\Mstardot/\Mstar,$ as a function of stellar mass and redshift, either for centrals or satellite galaxies.
\een
Specifically, we show results for the \um , \tng, and \galcus simulations using the in-situ star formation history, and show sSFR PDF results only for central galaxies.
To find the best-fit value, we minimize the summed mean squared error between all target PDFs and their corresponding predicted PDFs. In Appendix~\ref{diffstarpop_fitter_appendix}, we give a technical account of our probabilistic and differentiable techniques for fitting the parameters $\psisfh$ to the target PDFs, and show a wider range of comparisons for each simulation we compare to, including fits to the in-plus-ex-situ \um and \galcus runs (\S\ref{sec:exsitu}), sSFR fits for satellite galaxies for all five simulation runs (\S\ref{sec:satellites}), and some results on the best-fit relations predicted by \DstarPop in each simulation (\S\ref{sec:bestfit_diffpop}).

\begin{figure}
\includegraphics[width=\columnwidth]{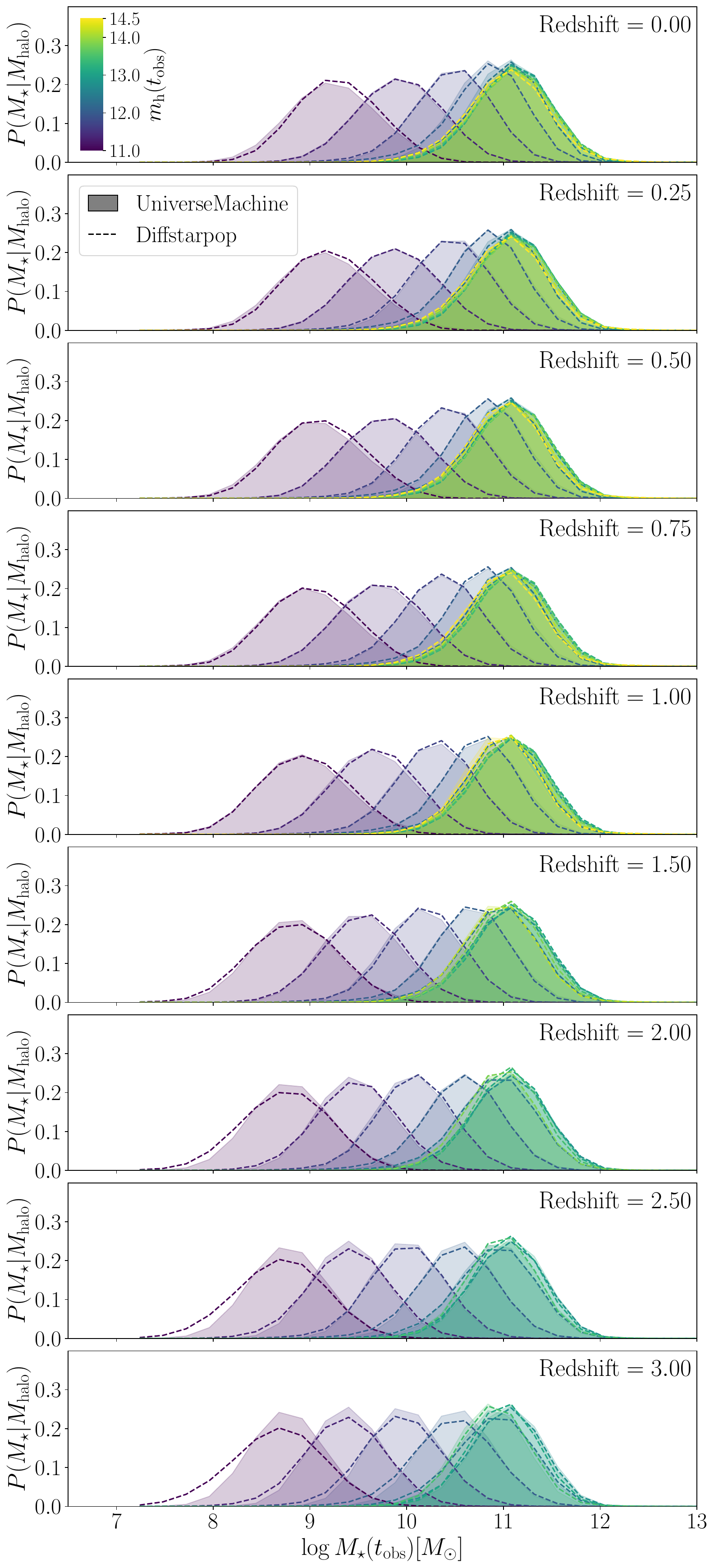}
\caption{{\bf \um stellar-to-halo-mass relation across redshift.} Each panel compares $P(\Mstar\vert\Mh, \zobs)$ in \um to its best-fitting \DstarPop counterpart. Different colored histograms show comparisons for different halo masses, as indicated in the legend. Shaded histograms show the stellar-to-halo-mass PDF in \um, and dashed lines show the best-fitting \DstarPop model.}
\label{fig_pdf_mstar_um}
\end{figure}

\begin{figure}
\includegraphics[width=\columnwidth]{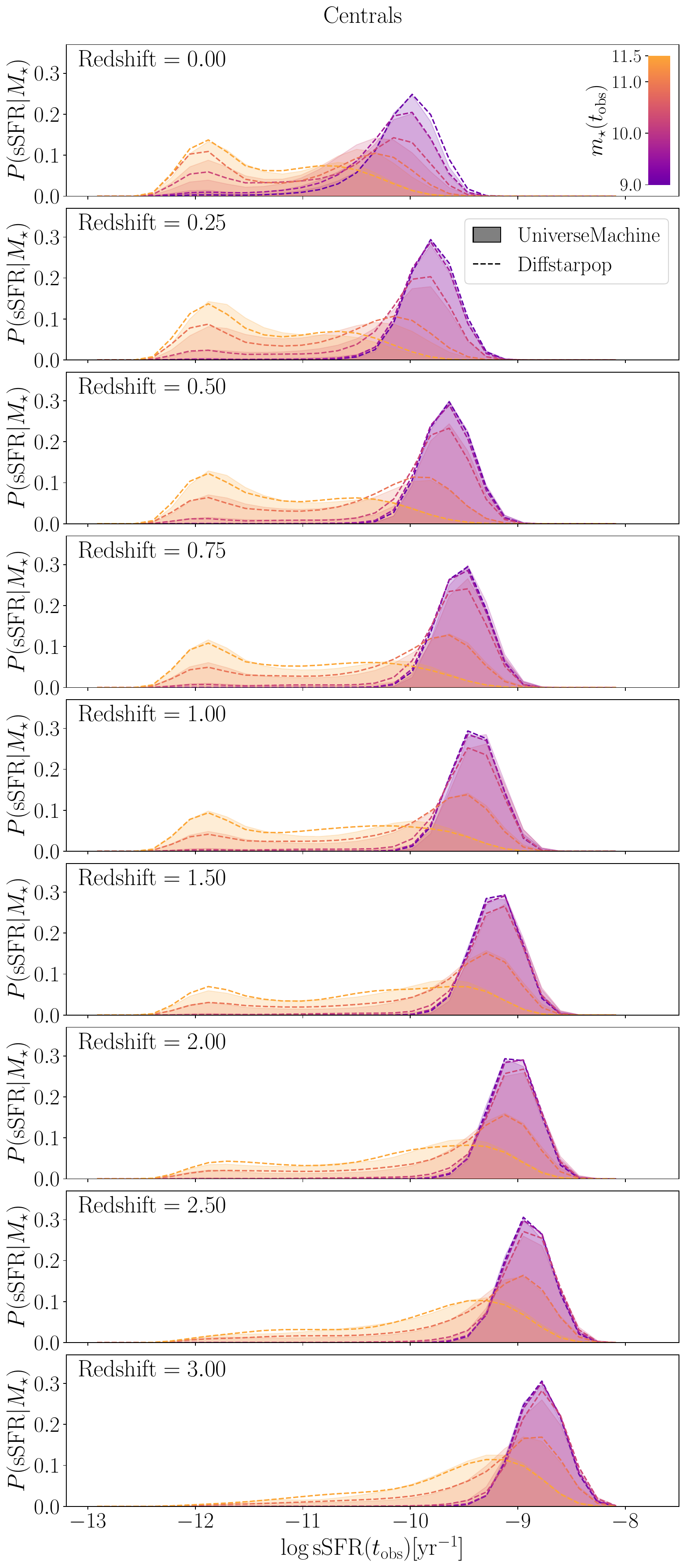}
\caption{{\bf \um sSFR PDF across redshift for centrals.} Similar to Fig.~\ref{fig_pdf_mstar_um}, but comparing $P({\rm sSFR}\vert\Mstar, \zobs)$ in \um to its best-fitting \DstarPop counterpart, only for central galaxies.}
\label{fig_ssfr_pdf_um}
\end{figure}

\begin{figure}
\includegraphics[width=\columnwidth]{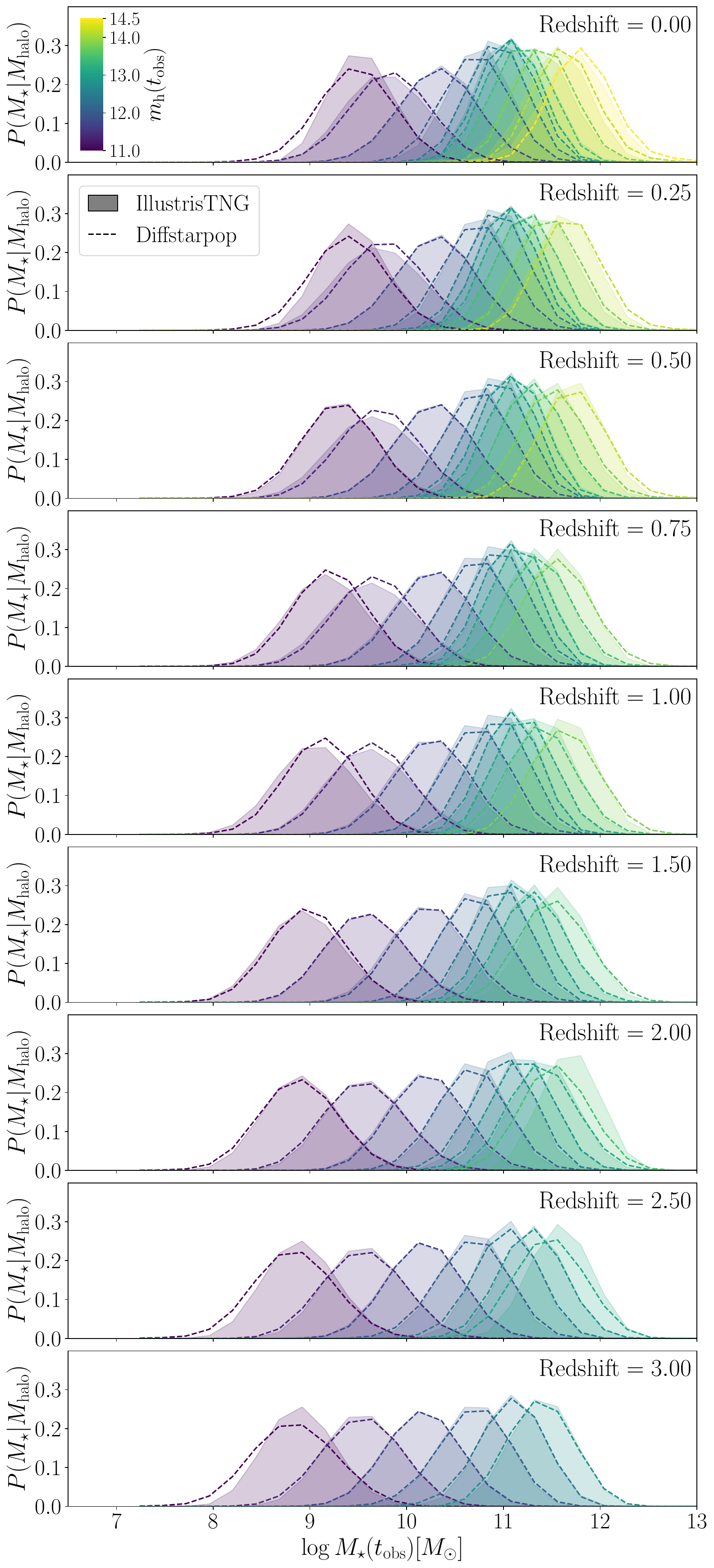}
\caption{{\bf \tng stellar-to-halo-mass relation across redshift.} Same as Fig.~\ref{fig_pdf_mstar_um}, for \tng.}
\label{fig_pdf_mstar_tng}
\end{figure}

\begin{figure}
\includegraphics[width=\columnwidth]{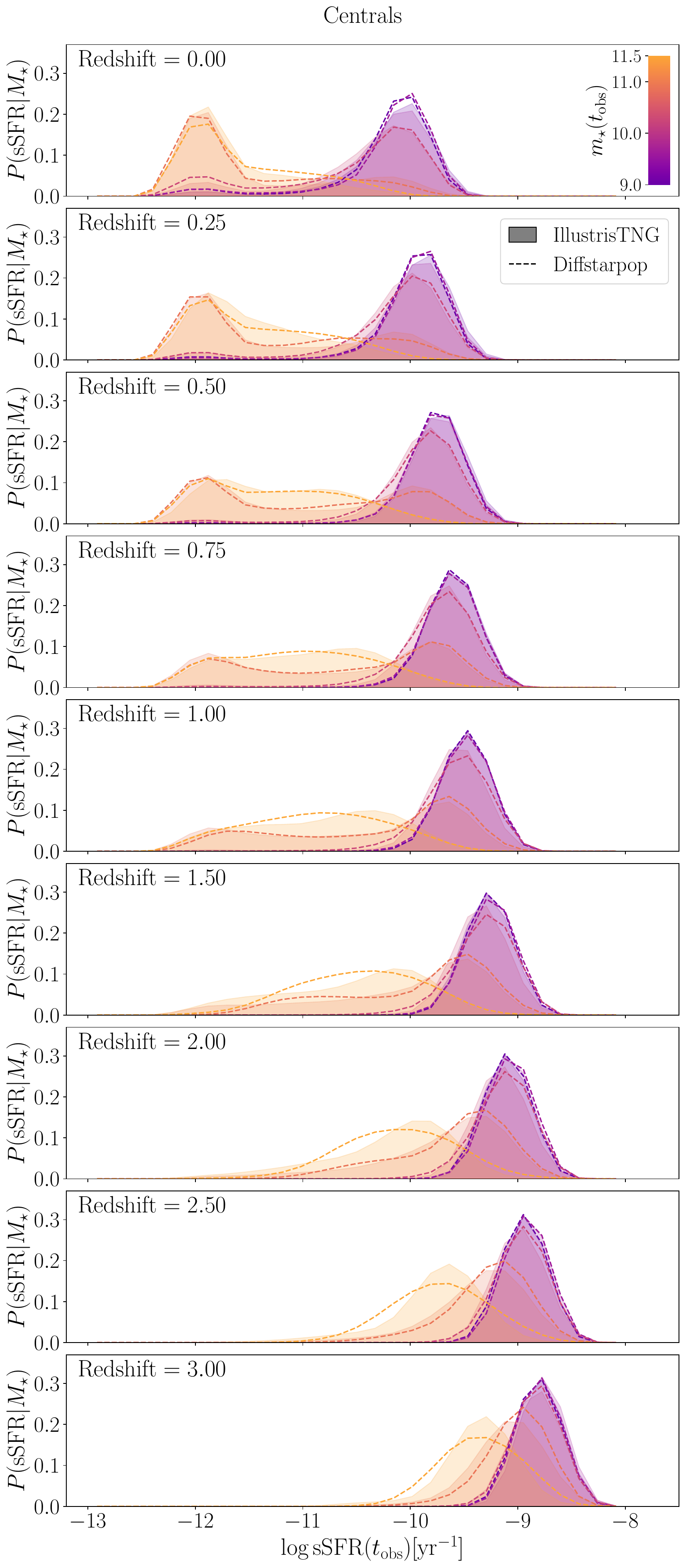}
\caption{{\bf \tng sSFR PDF across redshift for centrals.} Same as Fig.~\ref{fig_ssfr_pdf_um}, for \tng.}
\label{fig_ssfr_pdf_tng}
\end{figure}

In Figure~\ref{fig_pdf_mstar_um}, we show a comparison between \um and the best-fitting \DstarPop for the stellar-to-halo-mass PDF. The x-axis in each panel is stellar mass at each time/redshift $\Mstar(\tobs),$ and the y-axis is $P(\Mstar\vert\Mh,\zobs),$ the PDF of stellar mass conditioned on the value of halo mass at the observed redshift, $\Mh( \zobs);$ different panels show results for galaxies at different redshift. Within each panel, different colored histograms show comparisons for different halo masses, as indicated in the legend; shaded histograms show the stellar-to-halo-mass PDF in \um, and dashed lines show the best-fitting \DstarPop model. In Figure~\ref{fig_ssfr_pdf_um}, we show a similar comparison but for $P({\rm sSFR}\vert\Mstar, \zobs),$ where ${\rm sSFR}\equiv\log_{10}(\Mstardot/\Mstar),$ is the specific star formation rate at $\Mstar(\zobs)$.

\begin{figure}
\includegraphics[width=\columnwidth]{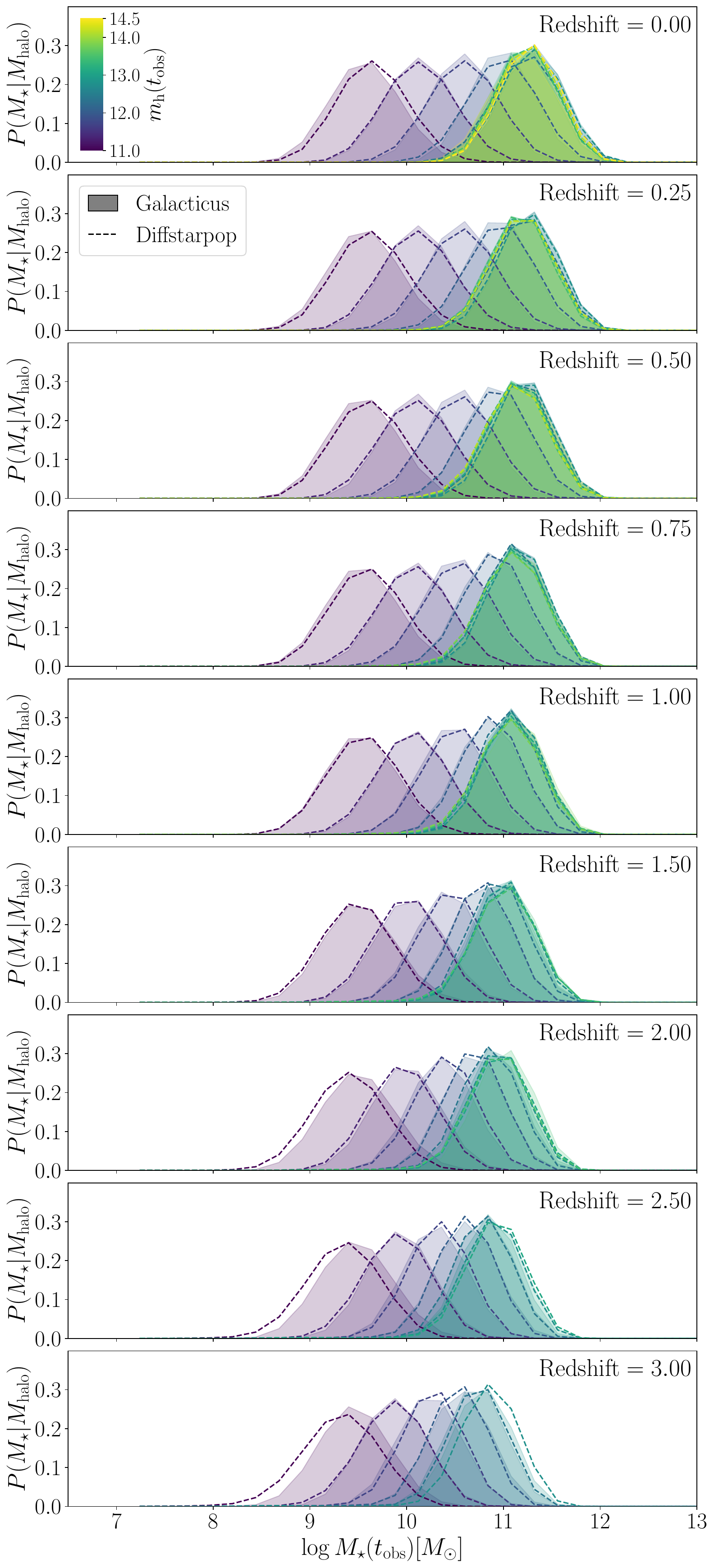}
\caption{{\bf \galcus (in-situ) stellar-to-halo-mass relation across redshift.} Same as Fig.~\ref{fig_pdf_mstar_um}, for \galcus.}
\label{fig_pdf_mstar_galcus}
\end{figure}

\begin{figure}
\includegraphics[width=\columnwidth]{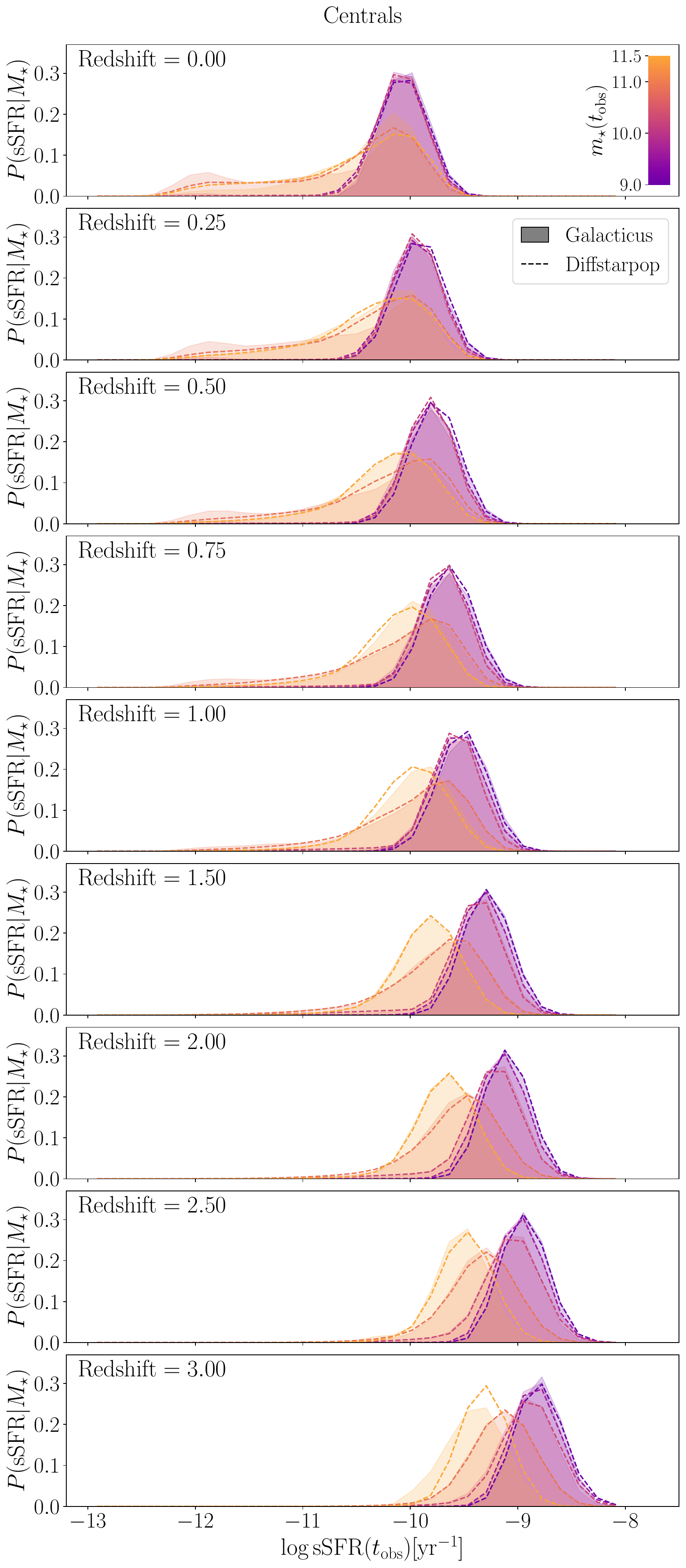}
\caption{{\bf \galcus (in-situ) sSFR PDF across redshift for centrals.} Same as Fig.~\ref{fig_ssfr_pdf_um}, for \galcus.}
\label{fig_ssfr_pdf_galcus}
\end{figure}

Figures~\ref{fig_pdf_mstar_um} \& \ref{fig_ssfr_pdf_um} show that \DstarPop is sufficiently flexible to approximate reasonably accurately the broad trends of how galaxy star formation connects to halo mass assembly, for halos of mass $\Mh(\tobs)=10^{11}\Msun-10^{14.5}\Msun$ hosting galaxies of mass $\Mstar(\tobs)=10^{9}\Msun-10^{11.5}\Msun$ in a broad redshift range spanning $0\leq z\leq 3.$ Figures~\ref{fig_pdf_mstar_tng} \& \ref{fig_ssfr_pdf_tng} show analogous results for \tng, and Figures~\ref{fig_pdf_mstar_galcus} \& \ref{fig_ssfr_pdf_galcus} for \galcus. Overall, we find that our formulation of \DstarPop is sufficiently flexible to approximate statistical distributions of galaxy SFHs seen in three very different simulated populations.

\section{Discussion}
\label{section:discussion}

In this paper, we have developed a new model for cosmological populations of galaxy star formation history (SFH), \DstarPop; the model can be used to paint galaxy SFHs onto merger trees in cosmological simulations. We have stress-tested the ability of \DstarPop to capture simulated populations of galaxy SFH predicted by the hydrodynamical simulation \tng, the semi-analytic model \galcus, and the semi-empirical model \um. We have focused in particular on validating our model based on its ability to faithfully reproduce the stellar-to-halo-mass PDF, $P(\Mstar\vert\Mh,z),$ and the ${\rm sSFR}-\Mstar$ PDF, $P({\rm sSFR}\vert\Mstar,z),$ because these conditional PDFs are closely linked to luminosity functions and color PDFs that are well-measured by current and near-future galaxy surveys \citep{Weaver2022,popcosmos_Deger,2025MNRAS.541..573D,2025A&A...699A.328A, COSMOS2025}. In Section~\ref{section:results} and Appendix~\ref{diffstarpop_fitter_appendix}, we showed that \DstarPop is able to reproduce each simulation's predictions for these distributions with a typical Kullback–Leibler divergence of 0.01–0.02 for the $\Mstar$ PDFs and 0.02–0.03 for the ${\rm sSFR}$ PDFs, and with an accurate recovery of both the mean and variance of these quantities across a wide range in halo mass and redshift. 

Using merger trees from $N$-body simulations, or cosmologically representative populations of dark matter halo assembly histories generated by DiffmahPop \citep{hearin_etal21_diffmah}, \DstarPop can produce self-consistent, cosmologically representative populations of galaxy star formation histories (SFHs). We emphasize that the diversity in these populations of MAHs and SFHs does not arise from adding random scatter around a mean relation. Instead, we explicitly parametrize the full statistical distributions of the \dmah and \dstar parameters that govern individual halo assembly and in-situ star formation rates. As a result, the model reproduces not only mean scaling relations, such as the stellar-to-halo mass relation, but also a cosmologically realistic diversity of individual halo and star formation trajectories.

\subsection{Towards Differentiable Sky predictions}

The smooth evolutionary trajectories of halo and galaxy growth in \DstarPop are not intended to reproduce very-short-term stochastic fluctuations in SFH \citep[e.g.][]{Tacchella2020, Iyer2020_burstiness, Iyer24_sfhvariability, Wang2022_stochastic}. When such SFH fluctuations occur long before the time of observation, there is a minimal impact on the statistical distributions of broadband galaxy colors \citep[e.g.][]{chaves_montero_hearin_2021_sbu2}; and such MAH fluctuations have only percent-level impact on halo clustering \citep{hearin_etal21_diffmah}. However, a large SFH fluctuation can indeed influence the spectrum of a galaxy when it occurs very close to the time of observation; in closely related work, we have incorporated this effect into the DSPS library using a new model for burstiness introduced in \citet{zacharegkas_etal25}.

In this paper, we have also neglected to directly model the role of major halo mergers, which can impact the SFH of satellite galaxies that eventually merge with their associated central. To account for these effects, we have developed Diffmerge, a model that incorporates ex-situ star formation into the \dstar framework. Diffmerge describes how the stellar mass formed in a satellite galaxy is deposited into its associated central, a crucial step for producing the most massive galaxies in the Universe. As shown in Appendix~\ref{diffstarpop_fitter_appendix}, \DstarPop can already approximate in- plus ex-situ SFHs through an effectively higher main-sequence efficiency. However, explicitly linking the additional quenching of satellite galaxies to their merging histories will be essential for accurately predicting the small-scale clustering of satellites \citep{moster_emerge1, behroozi_etal19}, and so we will explore this effect in follow-up work \citep{beltz_mohrmann_etal_inprep}.

Of course, \DstarPop is only an approximate model of the statistical distribution of galaxy SFH, and so cannot even in principle fully capture the complexity of SFHs in the galaxy populations in any of these simulations. However, our ultimate goal is not to create a statistical replica of these simulations; instead, with \DstarPop we aim to analyze the SFHs of observed galaxy samples. Observational measurements of the SED and photometry of a galaxy can only constrain a limited amount of information about its SFH \citep{ocvirk_etal06,carnall18_how_to_measure_sfh,leja18_how_to_measure_sfh,Lower2020,Iyer24_sfhvariability}, and so the residual errors in the ability of \DstarPop to approximate our target simulations may be subdominant compared to other sources of uncertainty in the problem, such as dust attenuation \citep[][]{Jones2022_dust,Jones23_dust,hahn_melchior_25_dust_biases_LSST}, or variations in the stellar spectral library, initial mass function (IMF), metallicity or burstiness prescriptions \citep{Haskell24_burstiness, Wang2024, tortorelli_etal25, Jones25_spsvariations, Bellstedt25_sedlibraries, Wang2025_sfhburstines}. In ongoing work, we are using \DstarPop in a larger pipeline, Diffsky: a simulation-based forward model of the SEDs and photometry of galaxy populations. Diffsky includes additional modeling ingredients that are needed to predict the SEDs of galaxy populations, such as a probabilistic model for dust attenuation curves, the model for short-timescale SFH fluctuations, i.e. burstiness, that we mentioned before, or the merging model that incorporates ex-situ effects. In follow-up work, we will use these additional models to quantify the success of \DstarPop in terms of its ability to analyze and interpret observations of the SEDs and photometry of galaxy populations measured by surveys such as the Vera C. Rubin Observatory Legacy Survey of Space and Time \citep[LSST,][]{ivezic_etal08}, the Nancy Grace Roman Telescope\footnote{\url{https://roman.gsfc.nasa.gov}}, and Euclid \citep{Euclid_paper1}. This forward-modeling framework can ultimately be extended to cosmological inference using cosmological evidence modeling techniques \citep{lange_etal19, lange_etal22} and differentiable large-scale structure summary statistics \citep{hearin_etal21_shamnet}.
Additionally, \DstarPop can be used as a prior on the \dstar parameters when fitting the SEDs of individual galaxies with SPS models. This approach allows us to derive Bayesian constraints on the physical properties of observed galaxies while incorporating assumptions about the physics of galaxy formation derived from complex models such as \um, \tng, or \galcus, or by constraints from observational data.

\subsection{Applications to generating Synthetic Galaxy Catalogs}

We have previously used a prototype version of \DstarPop to help create the publicly available synthetic galaxies in OpenUniverse2024 \citep{troxel_etal_openuniverse_2024}, a cross-collaboration effort between LSST and Roman. More recently, we used the model to populate the Discovery simulations \citep{gillian_discovery_sims_25}, tailored to investigate the way in which evolving dark energy impacts galaxy formation.
We have similarly used the model to help create mock galaxy catalogs for the DESI collaboration as part of the Emulator Mock Challenge \citep{beltz_mohrmann_etal_inprep2}. Our work in this area is comparable to closely related efforts in the field based on SAMs, which have been used for many years to generate synthetic galaxy catalogs \citep[e.g.,][]{merson_etal13,guo_etal13_Lgalaxies}. SAMs remain widely used to generate mocks for some of the most ambitious planned galaxy surveys; for example, the GAEA semi-analytic model is currently being used to create the Jiutian mocks \citep{han_etal25_jiutian_sims} for the China Space Survey Telescope (CSST) extra-galactic surveys \citep{cao_etal18_cssos,gong_etal19_cssos}, and in \citet{zhai_etal21_roman_galcus_mocks} \galcus was used to create a synthetic galaxy lightcone for the Roman survey telescope.

In SAMs, mock galaxies and their SFHs are painted onto simulated merger trees by numerically deriving a solution to a coupled ODE system for each individual halo. Recent work has begun to bring these SAM-based approaches into the differentiable-programming setting with \textsc{sapphire} \citep{Pandya206sapphire}, enabling local and global sensitivity analyses of galaxy state variables as well as gradient-based Bayesian inference. \DstarPop predictions are essentially parameterized families of solutions to these same ODE systems. For example, in \DstarPop we directly parameterize $\mseff(\Mh),$ whereas star formation efficiency and its scaling with halo properties is an {\em emergent} phenomenon in SAMs. This formulation greatly simplifies the task of implementing the model in a library of automatic differentiation (autodiff) such as JAX \citep{jax2018github}, which provides significant computational benefits. 
First, our fully parametric formulation guarantees that the exact same amount of computational work is vectorized across the galaxy population; this is precisely the setup in which GPUs achieve their greatest potential to accelerate scientific computing \citep[see, e.g.,][for an application to stellar population synthesis with 300-400x speedup on GPUs]{hearin_etal23_dsps}. 
Second, the availability of gradient information allows us to utilize advanced optimization algorithms to derive best-fitting model parameters; high-quality fits to our target PDFs are achievable with only a few hundred steps of gradient descent with modern learning rates. 
While in the present work we have only leveraged gradient information to optimize our model, gradient-based inference algorithms, e.g., Hamiltonian Monte Carlo \citep[HMC,][]{duane_etal87_hmc}, and the No U-Turn Sampler \citep[NUTS,][]{hoffman_gelman_2014_nuts} were recently used together with DSPS in \citet{zacharegkas_etal25} to derive Bayesian posteriors on the physical parameters of individual galaxies. Our future work will analogously deploy gradient-based algorithms to derive posteriors on the population-level parameters of \DstarPop.

\subsection{Relationship to Existing Forward-Modeling Approaches}

The recent literature includes several forward-modeling efforts that are closely related to the work presented here. In the GalSBI-SPS model \citep{Fischbacher25_SBI,tortorelli_etal25_galsbi}, analytical parameterizations are adopted so that galaxy redshifts and SPS properties can be sampled to generate populations. In PopSED \citep{li_etal24_popsed} and also in pop-cosmos \citep{alsing_etal24_pop_cosmos}, neural networks are used instead of analytical prescriptions to encode the form of the PDFs of SPS parameters. These population-level techniques have proven to be a powerful way to model cosmological redshift distributions \citep{Leistedt2023_popcosmos_photoz,thorp_alsing_etal24_pop_cosmos_photoz}, and study the evolution of star formation history \citep{popcosmos_Deger}. \DstarPop is a population-level model of galaxy SFH, not SPS properties, and so \DstarPop on its own does not have the predictive power of these models. We discuss below our ongoing use of \DstarPop as a key component in the Diffsky framework that additionally predicts SPS properties, but first we highlight a key distinction: \DstarPop is a model of the galaxy--halo connection, whereas dark matter halos are absent from the galaxy populations generated with the aforementioned population-level models. This formulation makes it straightforward to directly graft our model into simulated populations of halos, one of the principal applications of this work. For example, we are currently using \DstarPop to create galaxy populations with \tng-like, \um-like, and \galcus-like SFHs, and populate large-volume, high-resolution N-body simulations such as Last Journey \citep{heitmann_etal21_last_journey} with mock galaxy catalogs designed to support the analysis working groups in the DESI, LSST-DESC, and Roman collaborations. A straightforward extension of our current work would be to use the \DstarPop framework to populate N-body simulations with galaxies that mimic the SFH distributions in GalSBI-SPS, pop-cosmos, and related models, which we aim to pursue in future work.

Models of the galaxy--halo connection based on machine learning (ML) offer another efficient and differentiable approach to making predictions for cosmological populations of galaxies. There are by now many examples of such applications in the field. In \citet{paco_etal21_camels}, the authors introduced the CAMELS suite of hydrodynamical simulations that span a range of parameters in both cosmology and baryonic feedback; this suite has been used to train neural networks that can be used to make predictions for the galaxy population \citep[e.g.,][]{deSanti_etal23,hahn_etal24_cosmo_photometry_alone}. An analogous program based on semi-analytic models, CAMELS-SAM \citep{perez_etal23_camels_sam}, is also well underway. Much larger-volume hydro simulations such as \tng, EAGLE \citep{crain_etal15_eagle}, and Simba \citep{dave_etal19} have also been used to train neural networks to predict the galaxy population from an N-body simulation \citep[e.g.,][]{chen_mo_etal21_how_to_model_sfh,Chittenden_tojeiro_2023,das_etal24_ml_simba}.

AI-based models have much more flexibility than \DstarPop, since neural network architectures such as diffusion models or normalizing flows are essentially guaranteed to be able to accurately approximate almost any distribution \citep{tabak_vandeneijnden_2010_normalizing_flows,tabak_turner_2013_normalizing_flows}. This flexibility is a distinct advantage of AI-based models, which have demonstrated the ability to capture the detailed color PDFs exhibited by observed galaxy samples \citep[e.g.,][]{thorp_peiris_etal25_pop_cosmos2}. However, this flexibility comes at a cost: the neural networks used in population-level models typically have $N_{\rm dim}\gtrsim10^4$ or greater, which precludes the use of conventional Bayesian methods to quantify uncertainty on the best-fitting PDF \citep[although see][for a new approach to Bayesian sampling that makes this more practical]{behroozi25_ray_tracing}. By contrast, the formulation of \DstarPop presented here has 79 parameters, which is well within the capability of gradient-based samplers such as HMC and NUTS to quantify uncertainty with converged chains \citep[e.g.,][]{hoffman_gelman_2014_nuts}. This advantage also comes at a cost: the physical assumptions underlying our comparatively low dimensionality could lead to biased predictions due to our comparative inflexibility. To confront this concern, we have stress-tested the flexibility of our model using three very different galaxy simulations: \tng, \um, and \galcus. While we consider the results shown in \S\ref{section:results} and Appendix~\ref{diffstarpop_fitter_appendix} to be a promising indication that our model is suitably formulated, programmatically validating \DstarPop and each ingredient in the Diffsky framework will be a continued effort in future work.

\section{Conclusions}
\label{section:conclusions}

In this paper, we have introduced \DstarPop, a model for the galaxy--halo connection that makes predictions for the statistical connection between the star formation history (SFH) of galaxies and the mass assembly history (MAH) of their parent halos. We summarize our primary results below.
\begin{itemize}
    \item We have shown that \DstarPop is sufficiently flexible to capture the statistical distributions of galaxy SFHs of three very different simulations: \tng,  \galcus, and \um. In particular, \DstarPop is able to accurately recover the distribution of stellar masses, $P(\Mstar\vert\Mh, z),$ and the distribution of specific star formation rates, $P({\rm sSFR}\vert\Mstar, z),$ across a wide range of halo mass, $10^{11}\lesssim\Mh/\Msun\lesssim10^{14.5}$, stellar masses $10^{9}\lesssim\Mstar/\Msun\lesssim10^{11.5}$ and time, $0<z<3.$
    \item We have introduced a new version of \dstar, a model for individual galaxy SFH. The updated version simplifies the description of how accreted gas transforms into stars. As shown in Appendix~\ref{diffstar_appendix}, the updated model improves the description of stellar mass histories for \um, \tng and \galcus, and improves both the per-galaxy runtime and the memory efficiency by 10x. As a result, \DstarPop can generate $10^6$ galaxy SFHs in 1.1 CPU-seconds, or 0.03 GPU-seconds.
	\item We have introduced a new version of Diffmah, a model for individual halo MAH. The updated version includes a new parameter, $\tpeak,$ defined as the time at which the growth of the halo becomes arrested. As shown in Appendix~\ref{diffmah_appendix}, the updated model improves the description of halo assembly for subhalos with early infall times.
    \item \DstarPop is publicly available as part of the \Dstar code: \url{https://github.com/ArgonneCPAC/diffstar}. Our source code includes Monte Carlo generators that supply statistical samples of galaxy and halo assembly histories. 
\end{itemize}

\section*{Acknowledgements}

We thank Peter Behroozi, Benedikt Diemer, and Alexie Leauthaud for illuminating discussions. APH thanks The Invaders for {\em Spacing Out}. AA thanks The Strokes for \emph{Selfless}. We thank the developers of {\tt NumPy} \citep{numpy_ndarray}, {\tt SciPy} \citep{scipy}, Jupyter \citep{jupyter}, IPython \citep{ipython}, scikit-learn \citep{scikit_learn}, JAX \citep{jax2018github}, and Matplotlib \citep{matplotlib} for their extremely useful free software. While writing this paper we made extensive use of the Astrophysics Data Service (ADS) and {\tt arXiv} preprint repository.

We gratefully acknowledge use of the Bebop, Improv, and Swing supercomputers in the Laboratory Computing Resource Center at Argonne National Laboratory. Some of our computations were carried out with resources of the Argonne Leadership Computing Facility, a U.S. Department of Energy (DOE) Office of Science user facility at Argonne National Laboratory and is based on research supported by the U.S. DOE Office of Science Advanced Scientific Computing Research Program (ASCR).

The authors gratefully acknowledge the Gauss Centre for Supercomputing e.V. (www.gauss-centre.eu) and the Partnership for Advanced Supercomputing in Europe (PRACE, www.prace-ri.eu) for funding the MultiDark simulation project by providing computing time on the GCS Supercomputer SuperMUC at Leibniz Supercomputing Centre (LRZ, www.lrz.de). The Bolshoi simulations have been performed within the Bolshoi project of the University of California High-Performance AstroComputing Center (UC-HiPACC) and were run at the NASA Ames Research Center.

Work done at Argonne was supported under the DOE contract DE-AC02-06CH11357. A portion of this work was supported by the OpenUniverse effort, which is funded by NASA under JPL Contract Task 70-711320, ‘Maximizing Science Exploitation of Simulated Cosmological Survey Data Across Surveys’.

AA acknowledges support from project PID2023-153229NA-I00  funded by MICIU/AEI/10.13039/501100011033 and FEDER, UE. The project that gave rise to these results received the support of a fellowship from "la Caixa" Foundation (ID 100010434). The fellowship code is LCF/BQ/PI23/11970028.

\bibliographystyle{aasjournal}
\bibliography{bibliography}

\appendix
\counterwithin{figure}{section}

The material in these appendices provides more detailed information about \DstarPop than is contained in the main body, as well as more extensive comparisons of the model to target data from simulations. In Appendix~\ref{diffmah_appendix}, we discuss the new $\tpeak$ parameter of the Diffmah model introduced in this work. We show results from fitting the SFH of individual galaxies in simulations with the \dstar model in Appendix~\ref{diffstar_appendix}. In Appendix~\ref{diffstarpop_appendix}, we give a technical description of the mathematical formulation of \DstarPop. We describe our techniques for fitting the parameters of \DstarPop to various target PDFs in Appendix~\ref{diffstarpop_fitter_appendix}.

\vspace{0.3cm}

\renewcommand{\thefigure}{A\arabic{figure}}
\section{Diffmah fits to individual halo MAH}
\label{diffmah_appendix}

In this appendix, we discuss the role of the new Diffmah parameter $\tpeak$ introduced in this work. We refer the reader to \citet{hearin_etal21_diffmah} for a more comprehensive presentation of other aspects of the Diffmah model. For convenience, we begin with a brief overview of our methodology for deriving best-fitting parameters $\thetamah$ to the MAH of a simulated halo.

When fitting $\thetamah$ to simulated halos, we leverage the differentiable implementation of Diffmah and use the L-BFGS-B algorithm implemented in {\tt scipy} to minimize the mean logarithmic difference between the simulated and predicted values of $\Mh(t)$ \citep{broyden_1970_B_in_BFGS,fletcher_1970_F_in_BFGS,goldfarb_1970_G_in_BFGS,shanno_1970_S_in_BFGS}. When optimizing the parameters, we do not allow $\thetamah$ to take on arbitrary values, but instead we place physics-based priors on the parameter space. For example, a fundamental aspect of CDM structure growth is that dark matter halos experience a period of rapid growth at early times, and transition to a slow-accretion regime at late times \citep{bullock_etal01,wechsler_etal2002, tasitsiomi_kravtsov_2004}. When fitting Diffmah parameters to simulated MAHs, we therefore impose the following constraint: $$\aearly>\alate>0,$$ since we do not want the fitter to wander into an unphysical region of parameter space.

To implement these and other such constraints, we use invertible sigmoid functions $\Fbound$ to relate the physical Diffmah parameters to their unbounded counterparts, $\thetamah\equiv\Fbound(\uthetamah).$ In finding best-fitting Diffmah approximations, we run our gradient descents in the unbounded parameter space $\uthetamah,$ which ensures that the best-fitting $\thetamah$ will respect the parameter space constraints. We highlight this aspect of our fitting algorithm since we use this technique throughout this work.

\begin{figure}
\includegraphics[width=\columnwidth]{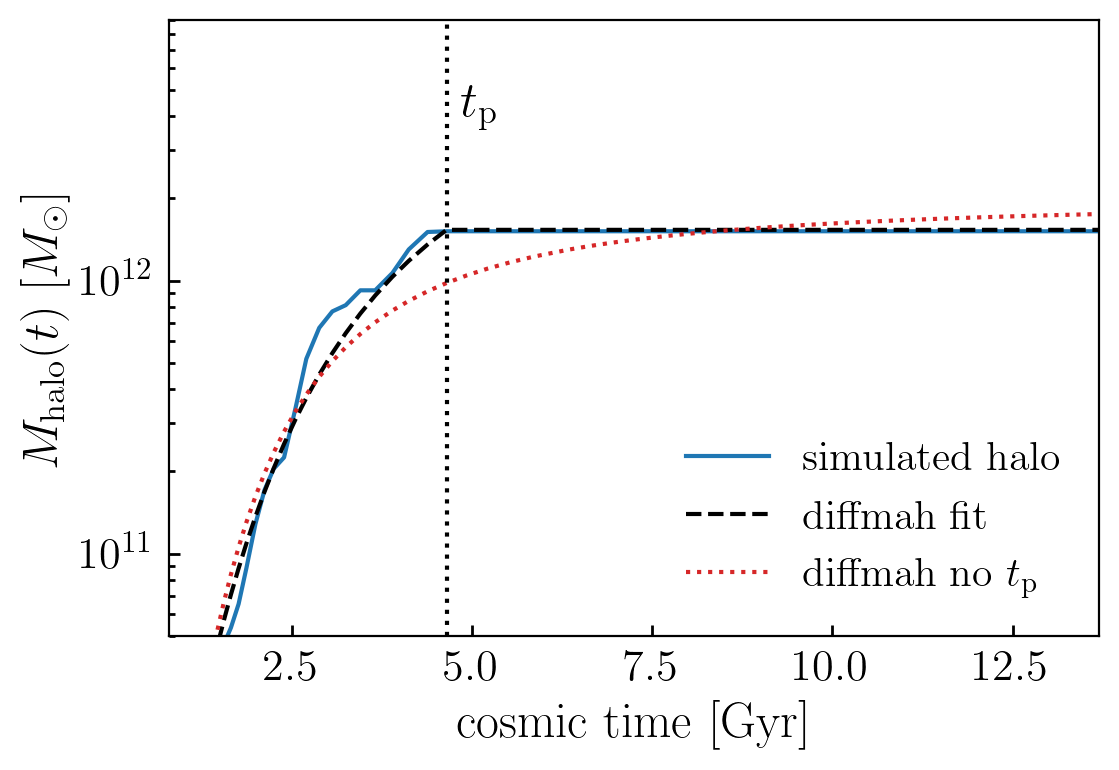}
\caption{{\bf Impact of new $\tpeak$ parameter on Diffmah fits.} The mass assembly history of an example subhalo in the SMDPL simulation is shown with the solid blue curve; the dashed black curve shows the best-fitting approximation of the updated Diffmah model introduced here; the dotted red curve shows the fit using the previously-published version of Diffmah, which does not include the $\tpeak$ parameter. The inclusion of the $\tpeak$ parameter generally improves the quality of the fits for subhalos with early infall times.}
\label{fig_diffmah_no_tpeak}
\end{figure}

\begin{figure}
\includegraphics[width=\columnwidth]{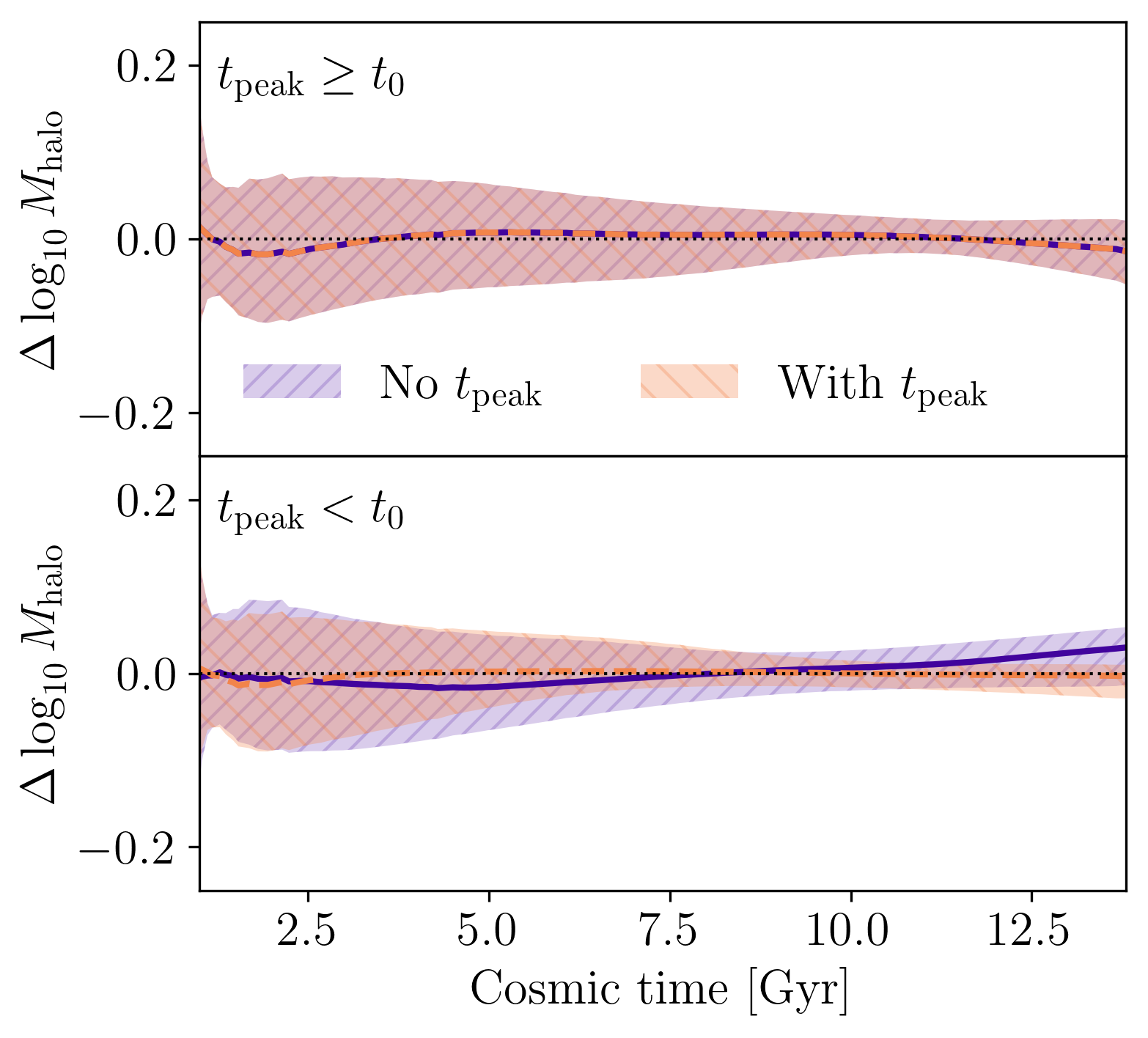}
\caption{{\bf Impact of new $\tpeak$ parameter on a population of Diffmah fits.} 
Residuals between the Diffmah model and the true halo mass accretion histories in \tng for haloes with $M_0>10^{11}M_{\odot}$. Solid lines show the median residual as a function of cosmic time, while the shaded regions indicate the 16th--84th percentile range across halos. Blue (solid) curves correspond to the original parametrization with no $t_{\rm peak}$, and orange (dashed) curves correspond to the updated model which include $t_{\rm peak}$. Halos are divided according to the best-fit value of $t_{\rm peak}$: the top panel shows halos with $t_{\rm peak} \ge t_0$, while the bottom panel shows halos with $t_{\rm peak} < t_0$. Including $t_{\rm peak}$ reduces both the bias and the scatter in the fitted mass accretion histories, and make no difference in haloes that did not need $t_{\rm peak}$.
}
\label{fig_diffmah_no_tpeak_tng}
\end{figure}

The effect of the $\tpeak$ parameter is defined in Eq.~\ref{eq_tpeak_def}, repeated below for convenience:
\beq
\nonumber
\Mpeak(t) = \begin{cases}
    \Mzero(t/t_0)^{\alpha(t)} & t\leq \tpeak , \\
    \Mzero(\tpeak/t_0)^{\alpha(\tpeak)} & t\geq\tpeak .
\end{cases}
\eeq
At times $t<\tpeak,$ the $\tpeak$ parameter has no effect on the MAH; for later times $t\geq\tpeak,$ $\Mpeak(t)$ is held constant to $\Mpeak(\tpeak).$

\begin{figure}
\includegraphics[width=\columnwidth]{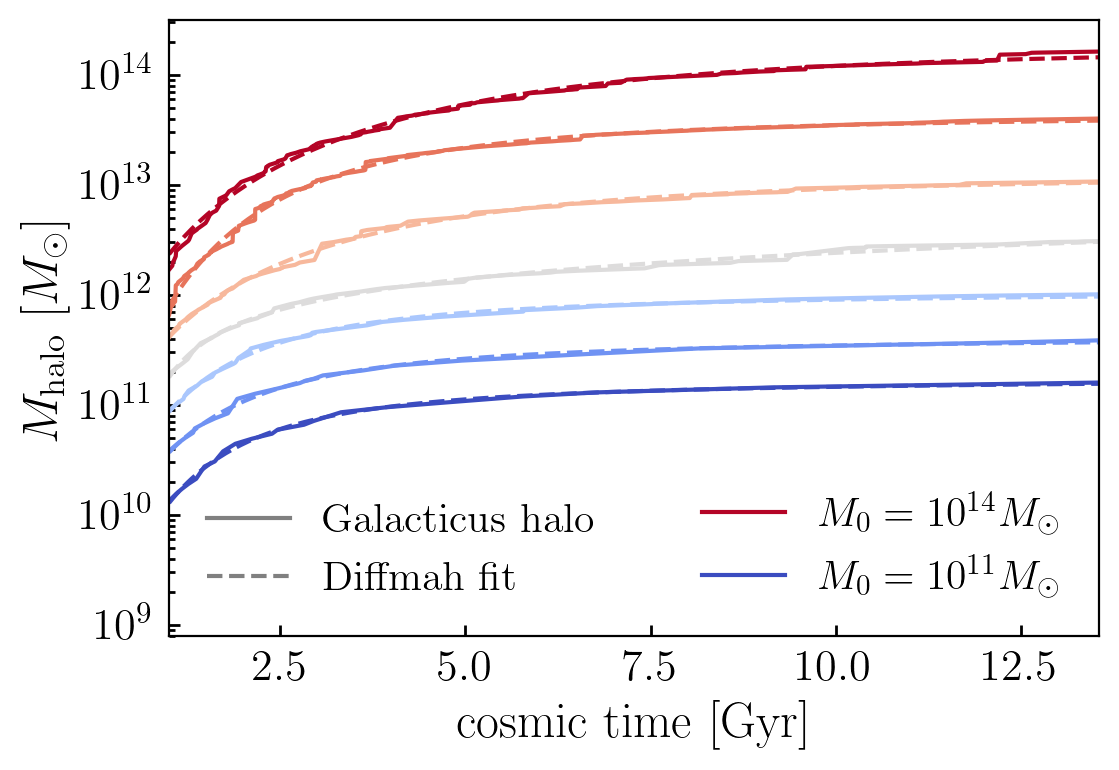}
\includegraphics[width=\columnwidth]{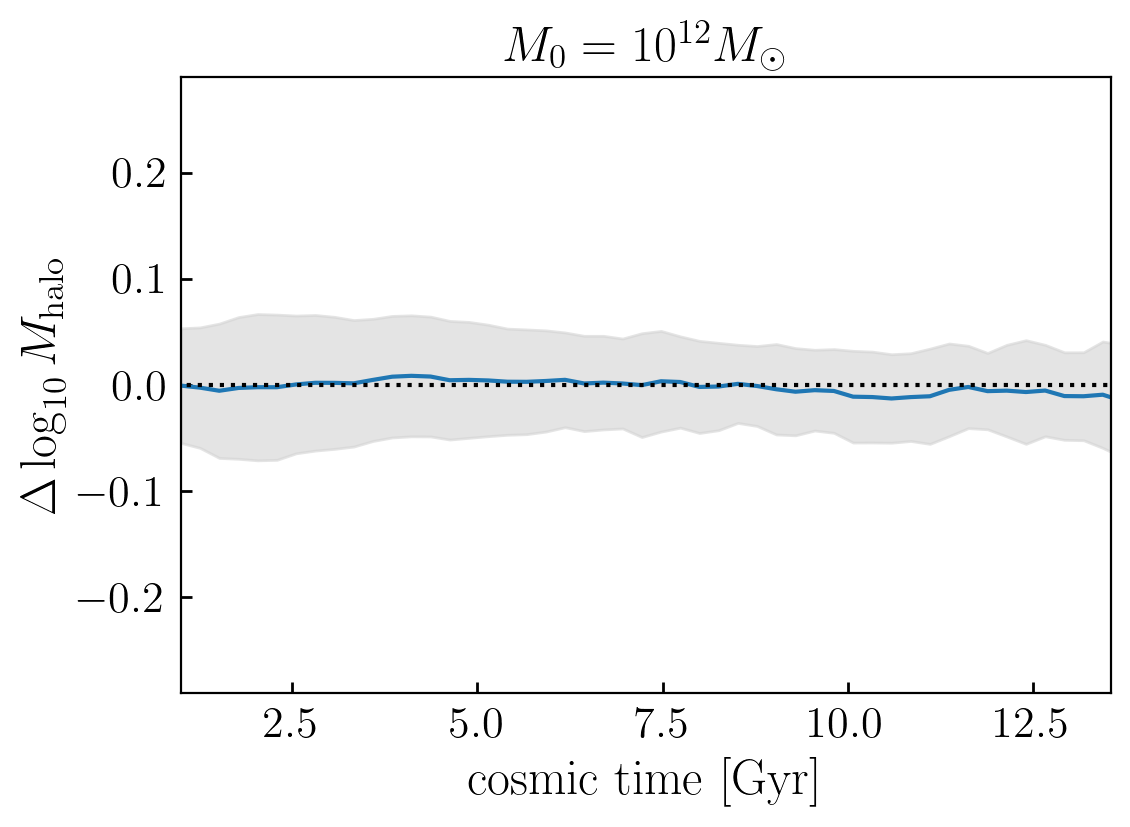}
\caption{{\bf Diffmah fits to Galacticus halos.} {\it Top panel:} Example Diffmah fits to selected individual Galacticus halos of different mass. The mass assembly history of a collection of halos is plotted as a function of time; different colored curves correspond to halos of different present-day mass, $M_0,$ as indicated in the legend; solid curves show the simulated MAHs, and dashed curves show the best-fitting Diffmah approximation. {\it Bottom panel:} Residual errors of Diffmah fits to a population of Galacticus halos with present-day mass $\Mhalo=10^{12}M_{\odot}.$ The average logarithmic difference between simulated and approximated halo mass is shown with the solid blue curve; the gray error band shows the variance of the logarithmic difference. Results for halos of different mass are similar.}
\label{fig_diffmah_galacticus_example}
\end{figure}

In Figure~\ref{fig_diffmah_no_tpeak}, we illustrate an example of a simulated subhalo in which $\tpeak$ plays a prominent role in the fit. On the y-axis is $\Mpeak$ plotted as a function of cosmic time on the x-axis; the solid blue curve shows $\Mpeak(t)$ taken directly from the simulation, the dashed black curve shows the best-fitting Diffmah model that includes the $\tpeak$ parameter, and the dotted red curve shows the best-fitting result of the original Diffmah model published in \citet{hearin_etal21_diffmah} that did not include $\tpeak.$ This subhalo has an especially early infall time into its host halo, and its growth was subsequently arrested by the strong tidal field within the host.

We can readily see in Figure~\ref{fig_diffmah_no_tpeak} that the dashed black curve gives a much better approximation to the simulated MAH relative to the dotted red, particularly around the time that the growth of the subhalo becomes arrested. We have visually examined hundreds of individual Diffmah fits, and we find this situation to be rather common, especially for massive subhalos with early infall times. When running fits to such halos using the original version of Diffmah, we find that the fitter is able to give a reasonable approximation such as the one shown with the dotted red curve in Figure~\ref{fig_diffmah_no_tpeak}, although the best-fitting values of $\thetamah$ tend to take on extreme values in order to compensate for the lack of parametric freedom to capture arrested growth. Meanwhile, when including $\tpeak$ in the model, the quality of the fits improve, and the remaining Diffmah parameters more closely resemble the regions of parameter space occupied by central halos. 
To quantify this improvement, in Figure~\ref{fig_diffmah_no_tpeak_tng} we show results for halos in the \tng simulation, finding similar results in the other simulations. For halos with $t_{\rm peak} < t_0$, introducing $t_{\rm peak}$ reduces both the bias and the scatter in the fitted mass accretion histories. In contrast, for halos with $t_{\rm peak} \ge t_0$, the two parametrizations yield nearly identical results. We take these observations as motivation to update the Diffmah model to include the $\tpeak$ parameter.

When fitting halo MAHs with Diffmah, we do not allow $\tpeak$ to vary, but instead we run the gradient descent with $\tpeak$ held fixed to the value calculated directly from the simulated MAH. We define this value as the first snapshot at which the maximum $\Mpeak$ is attained. When allowing $\tpeak$ to vary arbitrarily, we find that for a small outlier population of subhalos, the L-BFGS-B algorithm occasionally wanders into extreme regions of parameter space, especially for subhalos with only a small number of data points in the simulated MAH. Holding $\tpeak$ fixed to its simulated value ameliorates this situation. We have additionally explored an alternative version of our fitting algorithm in which $\tpeak$ is allowed to vary freely between the snapshots immediately before and after the directly-simulated value, finding no appreciable change in the results. This alternative may be preferable if continuously variable values of $\tpeak$ are preferred to the collection of discrete values defined by the simulation snapshots.

To generate our results, we have fit millions of simulated halos in each of our three simulations with Diffmah. 
In \citet{hearin_etal21_diffmah}, we used \um and \tng as our target simulations, and we showed that Diffmah is able to give an unbiased approximation of halo growth with a typical scatter of 0.1 dex for $t>1$ Gyr for all halos of present-day mass $M_{\rm halo}\gtrsim10^{11}M_{\odot}.$ As shown in Figure~\ref{fig_diffmah_no_tpeak_tng}, this overall level of scatter is maintained in the updated model with the addition of $\tpeak$, while the recovery of halos with $t_{\rm peak}<t_0$ is improved through a reduction in both bias and scatter. The present work is also the first time that Diffmah has been used to approximate halo assembly in Galacticus (i.e., as predicted by EPS merger trees), and so in Figure~\ref{fig_diffmah_galacticus_example} we give a visual demonstration of the quality of those fits. We find that Diffmah fits achieve the same level of accuracy when approximating the MAHs of Galacticus halos as when fitting halos in SMDPL and \tng.

\renewcommand{\thefigure}{B\arabic{figure}}
\section{Diffstar model of individual galaxy SFH}
\label{diffstar_appendix}

In \S\ref{subsec:diffstar_model}, we gave a high-level overview of the Diffstar model for individual galaxy SFH, and we showed an example fit to a simulated galaxy in \um to illustrate the basic behavior of the model. The Diffstar model was first introduced in \citet{alarcon_etal23_diffstar}, which contains a detailed pedagogical description of its ingredients. As noted in \S\ref{subsec:diffstar_model}, here we have presented an updated version of the model; however, its core physical principles remain the same. Specifically, \dstar is built on the following physical principles:

\begin{enumerate}
    \item A smooth parametric form (Diffmah) to describe dark matter halo growth.
    \item Baryonic (gas) accretion tracks dark matter accretion via a cosmic baryon fraction factor.
    \item Freshly accreted gas doesn’t convert instantly into stars. Instead, the transformation of gas into stars occurs gradually over time.
    \item Only a fraction of each accreted gas parcel ever transforms into stars.
    \item Galaxies may experience a quenching event that shuts down star formation, and quenched galaxies may subsequently rejuvenate and resume star formation.
\end{enumerate}

Motivated by results indicating the robustness of multi-Gyr gas depeletion timescales \citep{semenov_etal17_sfr_efficiency1}, in \citet{alarcon_etal23_diffstar}, the gradual conversion of gas into stars (step 3 above) was modeled via a parameterized function, $\mathcal{F}_{\rm cons}(t\vert\tau_{\rm cons}),$ the consumption function, whose shape was controlled by a free parameter, $\tau_{\rm cons}$, the gas consumption timescale. 
In that formulation, $\mathcal{F}_{\rm cons}(t\vert\tau_{\rm cons})$ explicitly describes how accreted gas becomes available for star formation over a timescale of $\tau_{\rm cons}$ through a convolution of the gas accretion history, $dM_{\rm g}/dt$. In practice, this consumption function must be flexible enough to capture the diversity of gas-processing behaviors found across different galaxy formation models and hydrodynamical simulations. Distinct physical prescriptions can potentially lead to markedly different gas depletion histories. If the functional form of the consumption kernel is too rigid, \dstar SFHs will not accurately reproduce the underlying trends of a given simulation or observed dataset. 

In our updated formulation of the \dstar\ model, we have removed this explicitly parameterized gas consumption function entirely. Instead, we directly compute the total accumulated baryonic mass available for star formation at each time, and quantify the efficiency with which gas is turned into stars as a timescale (with units of inverse time). As a result, variations in gas consumption arise naturally from the interplay between the evolving gas accretion history and the time-dependent efficiency, allowing recently accreted gas parcels to exhibit distinct consumption patterns from earlier-accreted material. We adopt the same parameterization for the scaling of efficiency with instantaneous halo mass as in \citet{alarcon_etal23_diffstar}. The efficiency peaks at a characteristic mass $\Mcrit$ ( $\sim10^{12}M_{\odot}$ for typical galaxies), and decreases monotonically at both lower and higher masses. 

A caveat of this approach is that gas already converted into stars is not removed from the available reservoir. Consequently, the inferred efficiency should be interpreted as an effective parameter that absorbs the impact of the missing depletion term, rather than as a directly literal measure of the instantaneous star-formation efficiency. This simplification reduces the physical interpretability of the efficiency parameter, but we adopt it as a practical tradeoff for computational efficiency, while also avoiding the need to impose a specific gas-consumption functional form that could be misspecified and bias the inferred star-formation histories.

\begin{figure}
\centering
\includegraphics[width=\columnwidth]{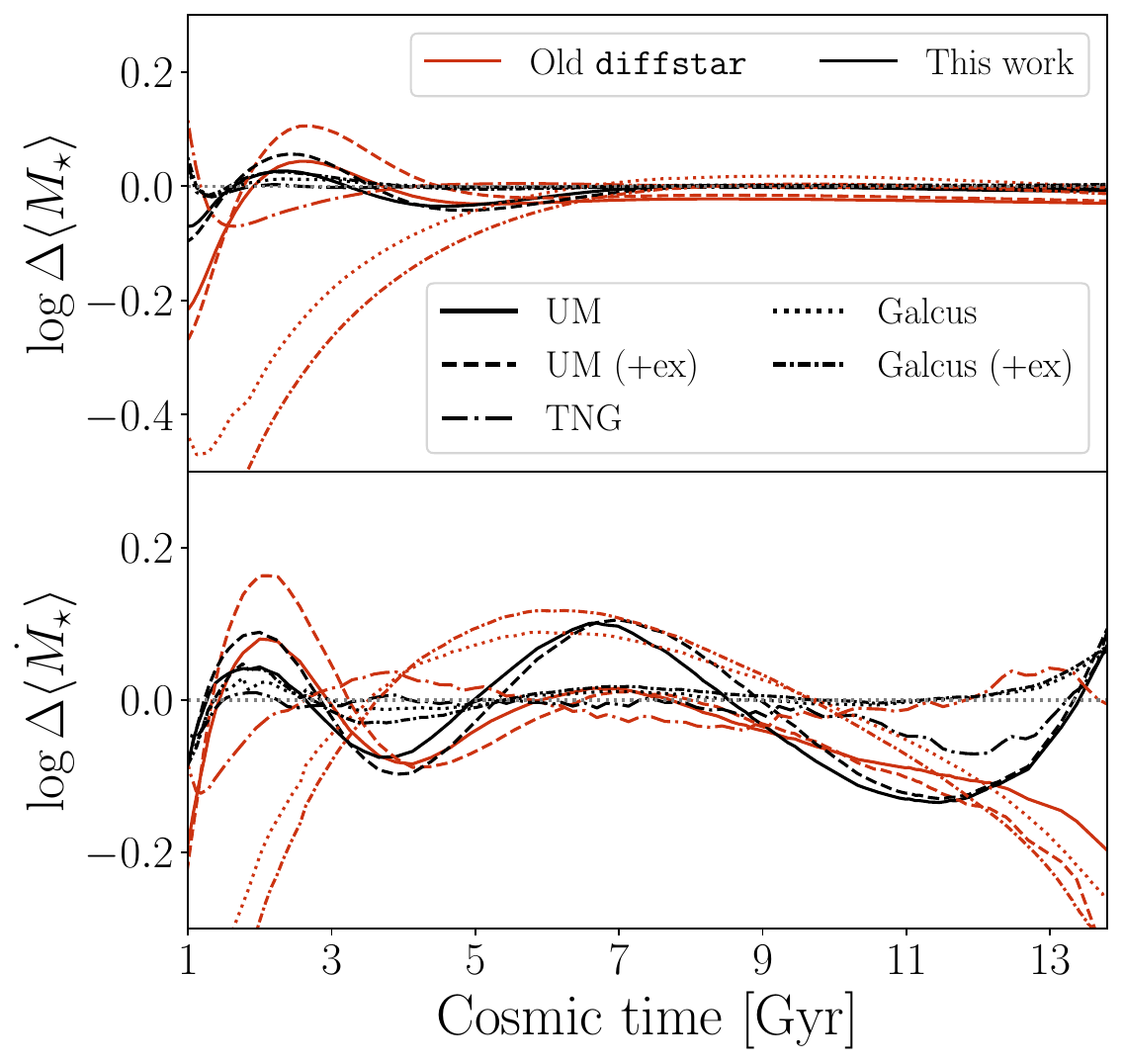}
\caption{{\bf Performance comparison of Diffstar} for halos with $\mpeakzero=12.5$, showing results from the previous model (red lines) and the updated model presented in this work (black lines) across five simulations (different line styles). The new formulation consistently yields lower bias in the residual logarithmic stellar mass and SFH differences and demonstrates improved accuracy within each simulation when comparing lines of the same style.
}
\label{fig_perfomance_diffstar_smh_sfh}
\end{figure}

\begin{figure}
\centering
\includegraphics[width=\columnwidth]{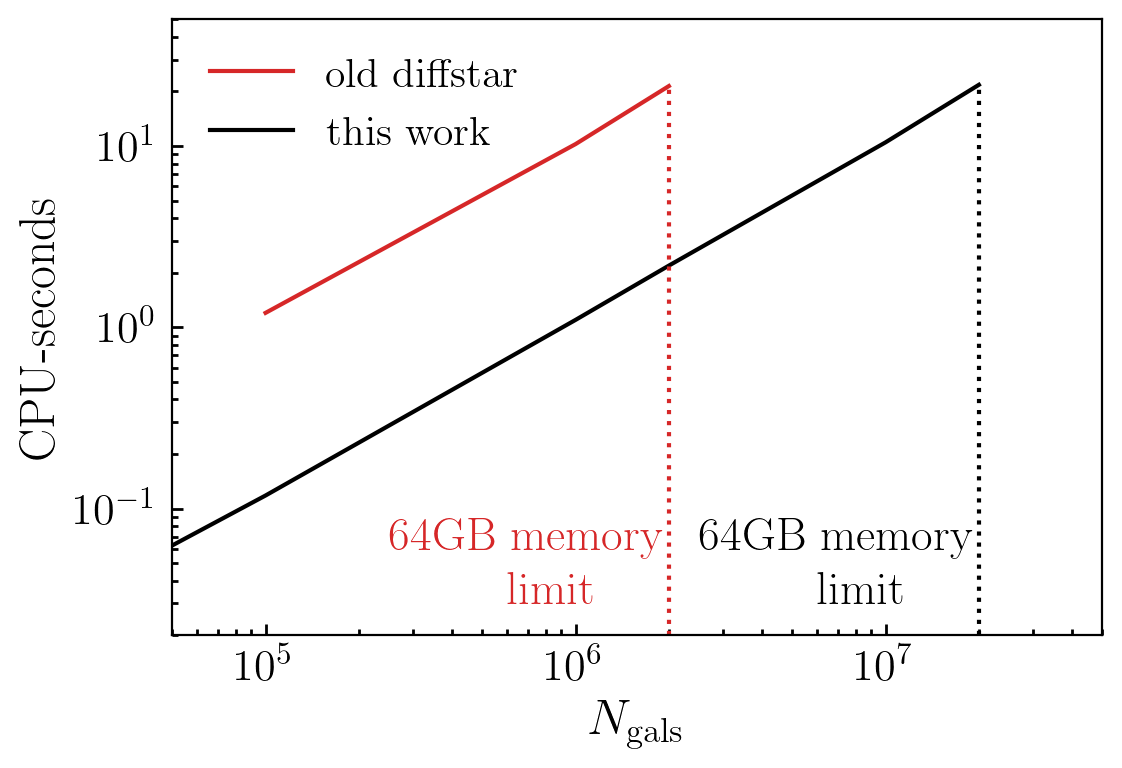}
\caption{{\bf Performance comparison of Diffstar}. The figure shows the runtime (in CPU-seconds) required to calculate SFHs for a galaxy population of size indicated by the x-axis. The dotted vertical lines indicate the number of galaxies at which each version exceeds 64 GB of system memory.
}
\label{fig_perfomance_diffstar_time_memory}
\end{figure}

These updates to the \dstar\ formulation provide a simpler and more efficient foundation for modeling galaxy SFHs while preserving the model’s physical interpretability. The new approach no longer requires specifying an uncertain gas consumption function and achieves higher accuracy in reproducing stellar mass and star formation histories across \um, \tng, and \galcus, as shown in Figure~\ref{fig_perfomance_diffstar_smh_sfh}. In addition, removing the convolution step yields more than an order-of-magnitude improvement in both runtime and memory efficiency (Figure~\ref{fig_perfomance_diffstar_time_memory}). These advances are particularly important for \DstarPop, whose success depends critically on accurate fits to individual galaxy SFHs. In the following, we validate the updated Diffstar model and demonstrate that it achieves high-quality fits across multiple simulations.

In \citet{alarcon_etal23_diffstar}, the in-situ SFHs of millions of galaxies simulated in \tng and \um were fit with Diffstar, and it was shown that the model is sufficiently flexible to describe the average stellar mass histories of galaxies in both simulations with an accuracy of $\sim0.1$ dex across most of cosmic time for a wide range of galaxy masses. The first step in deriving a Diffstar fit is approximating the halo MAH with Diffmah, and in this work we have used a new version of Diffmah that includes the $\tpeak$ parameter (see Appendix~\ref{diffmah_appendix}). We have therefore repeated the Diffstar analysis in \citet{alarcon_etal23_diffstar} for \um and \tng using the updated models of Diffmah and Diffstar. Additionally, we have subjected Diffstar to a new stress-test of its flexibility by using SFHs in the \galcus semi-analytic model as target data, and also by using SFHs that include ex-situ components for both \um and \galcus. 

\begin{figure*}
\centering
\includegraphics[width=\textwidth]{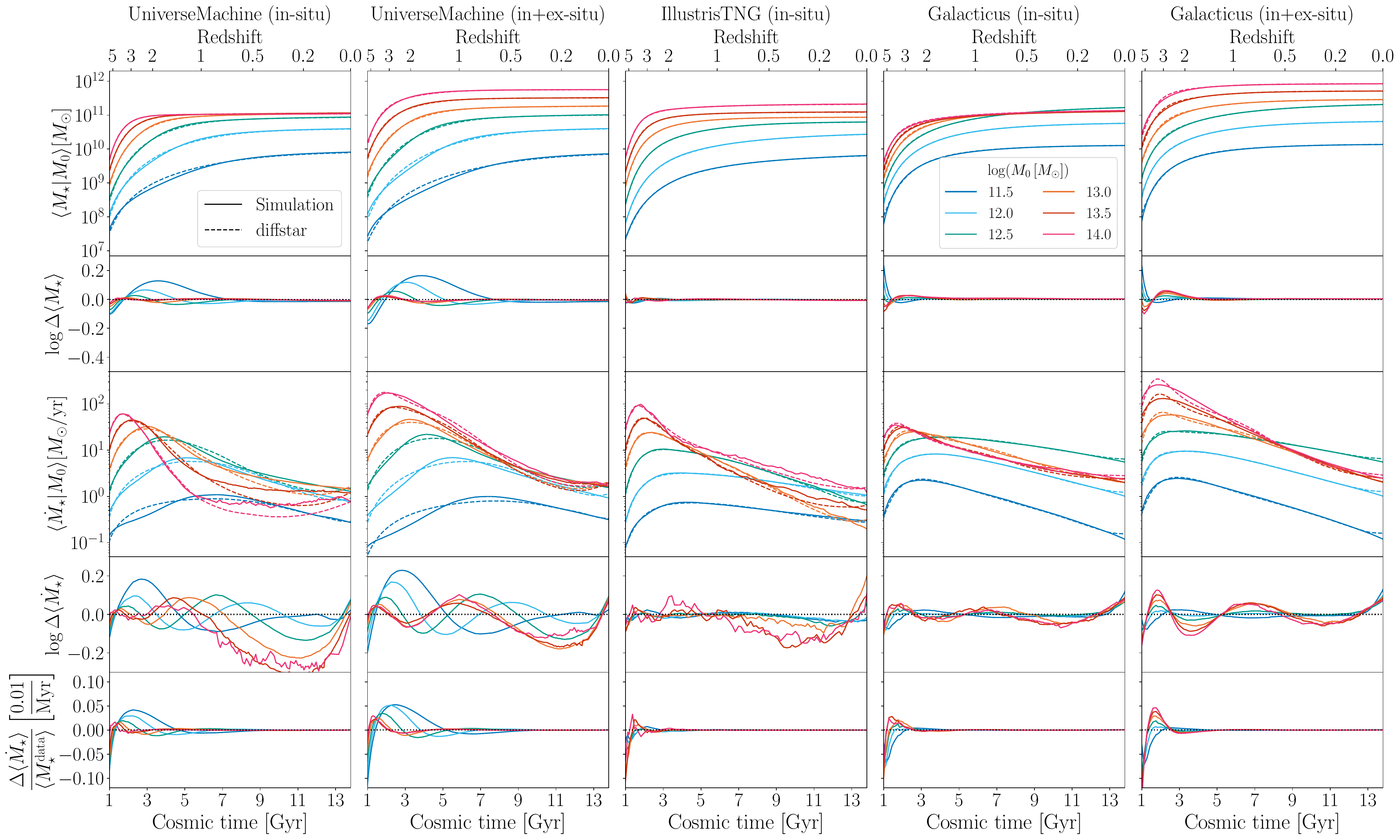}
\caption{{\bf Quality of Diffstar fits to individual galaxy SFH}. 
\textit{1st row of panels:} Average stellar mass histories (SMH) as a function of cosmic time for halos of different present day mass $M_0\equiv\Mpeak(z=0)$. Dashed curves show the prediction we obtain by fitting Diffstar to each individual SMH history and taking the average for the same halos.  \textit{2nd row:} Residual logarithmic difference between the average SMH in the simulation and the average prediction from Diffstar. \textit{3rd row}: Star formation histories (SFH). \textit{4th row}: Residual logarithmic SFH difference. \textit{5th row}: Residual SFH difference normalized by the simulation SMH in  $[(100\,{\rm Myr})^{-1}]$ units. The different columns show results for each of the five simulation runs studied in this work (see labels at the top).
}
\label{fig_average_Mstar}
\end{figure*}

Figure \ref{fig_average_Mstar} demonstrates the high fidelity of the updated \dstar\ model in reproducing the stellar mass and star formation histories (SMHs and SFHs) across a diverse suite of simulations. For all five runs, \um\ (in-situ and in+ex-situ), \tng, and \galcus\ (in-situ and in+ex-situ), the top panels show that the Diffstar-predicted average SMHs (dashed curves) closely follow the simulation results (solid lines) over the full range of halo masses and cosmic times. The residual panels indicate that the typical deviations remain below $\approx$0.1 dex for most of cosmic history, confirming that the model accurately captures both the normalization and shape of the growth curves. The largest discrepancies occur at early times, particularly for \um\ at $z\sim2$ and $\mpeakzero = 11.5$, where halos are resolved with fewer than 200 SMDPL particles; this suggests that limited early-time halo resolution may contribute to the mismatch. For comparison, \tng\ has more than an order of magnitude higher particle resolution in this regime. By contrast, the agreement is uniformly strong at later times and for larger halo masses across all simulations, suggesting that the model is especially well suited to applications focused on low-to-intermediate-redshift stellar assembly, while some caution is warranted when interpreting regimes affected by limited particle resolution. 

The corresponding SFHs exhibit similarly small residuals, and the last row puts those residuals differences in perspective by dividing them by the stellar mass history (similar to an error in specific star formation rate). These residuals from the last row are in units of inverse 100\,Myr, and they can be directly compared to those presented in Figure~5 from \cite{Lower2020} for ``non-parametric" and delayed$-\tau$ SFH models. Even in the stress-test cases including ex-situ stellar components, Diffstar maintains sub-0.1 dex accuracy in SMH, demonstrating its flexibility in adapting to different galaxy formation prescriptions. Overall, this figure validates that the combination of the revised Diffmah and Diffstar models provides a robust, computationally efficient, and broadly applicable description of galaxy growth across hydrodynamical, semi-analytic and empirical models of galaxy formation.

\renewcommand{\thefigure}{C\arabic{figure}}
\section{DiffstarPop model formulation}
\label{diffstarpop_appendix}

This appendix contains a technical description of the formulation of \DstarPop. As specified in Section~\ref{section:model_form}, \DstarPop is a parametric model for $P(\thetasfh\vert\thetamah),$ the probability that a dark matter halo with MAH given by $\thetamah$ hosts a galaxy with SFH given by $\thetasfh$. The \dstar parameters are based on ingredients of galaxy formation, and therefore they are bounded by limits that enforce a physical galaxy SFH. These limits are enforced by a set of sigmoid-bounding functions, $\mathcal{B}(x)$, that relate the bounded, physical diffstar parameters $\thetasfh$, to an unbounded space, $\uthetasfh$, i.e. $\thetasfh = \mathcal{B}(\uthetasfh)$, which is analytically invertible. For details see the \Dstar repository. The \DstarPop model is defined in the unbounded space of Diffstar parameters, $\uthetasfh$, as a Gaussian mixture model of two components, consisting of a ``quenched" population of galaxies that experiences a significant quenching event before today ($\qtime<t_0$), and a ``main sequence" population that never experiences such quenching,
\begin{equation} \label{eq:diffstarpop_definition}
\begin{split}
    P(\uthetasfh\vert\thetamah, \psisfh) =& (1-\fquench)\, P_{\rm ms} + \fquench \, P_{\rm q};\\[0.5em] 
    \text{with:} \qquad P_{\rm ms} =& P_{\rm ms}(\uthetasfh\vert\thetamah, \psisfhms); \\
    P_{\rm q} =& P_{\rm q}(\uthetasfh\vert\thetamah, \psisfhq);\\
    \fquench=&\fquench(\psisfhfquench).
\end{split}
\end{equation}
The \DstarPop parameters $\psisfh=\{\psisfhms, \psisfhq,\psisfhfquench\}$ specify how the mean and covariance of the Gaussian distributions $P_{\rm ms}$ and $P_{\rm q}$, as well as the mixing parameter (the quenched fraction $\fquench$), depend on the halo properties $\thetamah$.

To develop an intuition for each scaling relation, we first fit the SFHs and MAHs of millions of simulated galaxies and their corresponding host halos with \dstar and \dmah, obtaining their best-fit $\thetasfh$ and $\thetamah$ values, respectively. We then studied how the mean and standard deviation of each parameter scaled with present-day halo mass $\mpeakzero$, both for quenched ($\qtime<t_0$) and main sequence ($\qtime>t_0$) galaxies, where $\qtime$ is the Diffstar quenching time parameter, and $t_0$ the present-day. Once the scaling relations were developed, we roughly calibrated them against these individual SFH fits, and we used those as our initial guess for the parameter calibration discussed in Appendix~\ref{diffstarpop_fitter_appendix} and shown as part of our main results in Section~\ref{section:results}. 

The mean and standard deviation of the Diffstar parameters, for both quenched and main sequence galaxies, approximately follow a linear relation with $\mpeakzero$. To capture this trend, we model each with two free parameters: a slope and an intercept evaluated at $\mpeakzero = 12.5$. To prevent the linear model from producing unphysical or mathematically invalid values (such as a negative standard deviation), we use a ``smoothly clipped line" model, $\sclip(x)$. This approach retains the linear form within a valid range but smoothly limits the output by imposing fixed minimum and maximum values on the mean and standard deviation, with bounds that are set a priori.

For most \dstar parameters, the mean scaling relations $\langle \cdot \mid \mpeakzero \rangle$ are well described by a clipped linear dependence on halo mass. However, in our simulations we find that three parameters, $M_{\rm crit}$, $\epsilon_{\rm crit}$, and the quenching time $\qtime$, are better represented by a model in which the slope of the relation changes smoothly with $\mpeakzero$. For example, the relation for $M_{\rm crit}$ increases with halo mass at low $\mpeakzero$ and then flattens at higher $\mpeakzero$, behavior that is not well captured by a single clipped linear model. To describe these trends, we use a four-parameter ``sigmoid-slope" (sigslope) model, in which the slope of the relation transitions smoothly between two asymptotic values. This model both matches the forms seen in the simulations and yields improved fits in \DstarPop\ compared to a clipped linear model. Figure~\ref{fig:diffstarpop_mean_ecrit} shows an example of this behavior for $\langle \log \epsilon_{\rm crit} \mid \mpeakzero \rangle$ measured in \um, \tng, and \galcus, along with the corresponding best-fit sigslope relations from \DstarPop. The same sigslope parameterization is used for $M_{\rm crit}$ and for $\qtime$, and it can be described as:

\begin{equation}
    \langle\Mcrit \vert \mpeakzero \rangle = b + S(\mpeakzero, x_0, k, m_{\rm lo}, m_{\rm hi})\cdot (\mpeakzero - x_0)
\end{equation}
where $S(x)$ is a sigmoid as in Eq.~\ref{eq:sigmoid}, we fix $k=3$, and $\{b, x_0, m_{\rm lo}, m_{\rm hi}\}$ are the four free parameters of this model, with $b$ being the intercept at $x_0$, the latter being also the value where the sigmoid transitions from slope $ m_{\rm lo}$ to $ m_{\rm hi}$.

Overall, to model the means and standard deviations of the quenched (8 diffstar parameters) and main sequence (4 parameters) subpopulations of all of these parameters, we end up introducing a total of 58 \DstarPop $\psisfh$ parameters that depend only on $\mpeakzero$. In the next subsection, we describe the model for the quenched fraction and show that the quenching time further depends on the halo formation time, characterized by $\tpeak$.

\subsection{Quenching and dependence on halo formation time} \label{sec:subsec_tpeak}

\begin{figure}
\includegraphics[width=\columnwidth]{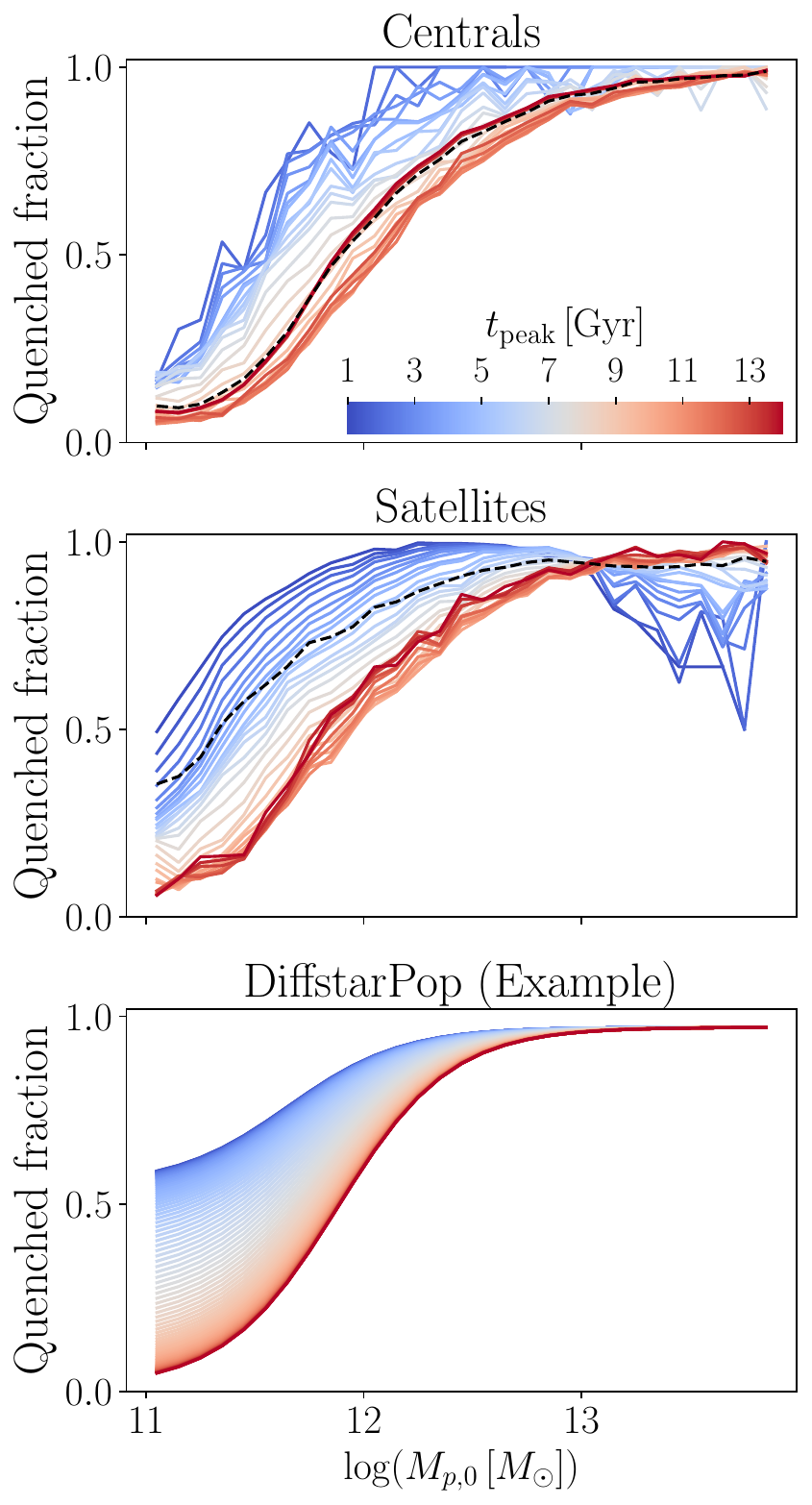}
\caption{{\bf Quenched fraction dependence on halo mass and $\tpeak.$} The panels show the quenched fraction of central (top) and satellite (middle) galaxies in \um as functions of the present-day peak halo mass $\mpeakzero$ and of the Diffmah parameter $\tpeak$ (different color lines). The black dashed lines indicate the quenched fraction averaged over $\tpeak$. The bottom panel illustrates an example of the \DstarPop model, highlighting how the quenched fraction varies smoothly with both $\mpeakzero$ and $\tpeak$.}
\label{fig:fquench_tpeak}
\end{figure}

The quenched fraction represents the proportion of galaxies that have experienced a significant quenching event before the present time, $\qtime<t_0$ (see Eq.~\ref{eq:diffstarpop_definition}).
We model this quantity as a sigmoid function of $\mpeakzero$, with four free parameters for central galaxies and another four for satellite galaxies.  Figure~\ref{fig:diffstarpop_mean_qfrac} compares the quenched fractions measured from \um, \tng and \galcus, along with the best-fit sigmoid approximation from \DstarPop. 

From fitting thousands of individual halos, we find that the quenching time depends not only on the present-day halo mass but also on the halo formation time. To capture this secondary dependence, we introduce an explicit dependence on the Diffmah parameter $\tpeak$ within \DstarPop.
Specifically, we allow two of the sigmoid parameters, ($y_{\rm lo}$ and $x_0$) to themselves vary smoothly with $\tpeak$ according to additional sigmoid relations. Overall, the quenched fraction for either central or satellite galaxies can be described as
\begin{equation}\label{eq:fquench}
\begin{split}
    \fquench(\mpeakzero \vert \psisfhfquench) &= S(\mpeakzero, x_0, k, f_{\rm lo}, f_{\rm hi}) , \\[0.5em]
    \text{with:}\qquad x_0 &= S(\tpeak, \tilde{x}_0, \tilde{k}, y^{x_0}_{\rm lo}, y^{x_0}_{\rm hi}) , \\
    f_{\rm lo} &= S(\tpeak, \tilde{x}_0, \tilde{k}, y^{f_{\rm lo}}_{\rm lo}, y^{f_{\rm lo}}_{\rm hi}) . \\
\end{split}
\end{equation}
For quenched fraction of centrals galaxies there are a total of 8 free parameters: $\psi_{\rm SFH}^{\rm fqch, cen} = \{ x_0, k, \tilde{x}_0, \tilde{k}, y^{x_0}_{\rm lo}, y^{x_0}_{\rm hi}, y^{f_{\rm lo}}_{\rm lo}, y^{f_{\rm lo}}_{\rm hi} \}$, and similarly there are another 8 parameters for satellite galaxies.

Figure~\ref{fig:fquench_tpeak} illustrates the flexibility of our quenched fraction model. The top and middle panels show the measured quenched fraction in the \um simulation (only in-situ) for central and satellite galaxies, as a function of both $\mpeakzero$ and $\tpeak$. The bottom panel shows an example \DstarPop prediction for the same quantity. Different line colors correspond to different values of $\tpeak$, while the black dashed lines indicate the quenched fraction averaged over $\tpeak$. In the top panel, the dashed line lies close to the $\tpeak\sim t_0$ curves, since most central galaxies have $\tpeak\geq t_0$, while the distribution of $\tpeak$ for satellites is much broader, especially at lower halo masses ($\mpeakzero\lesssim12.5$), so the dashed line appears at significantly earlier $\tpeak$ values.

\begin{figure}
\includegraphics[width=\columnwidth]{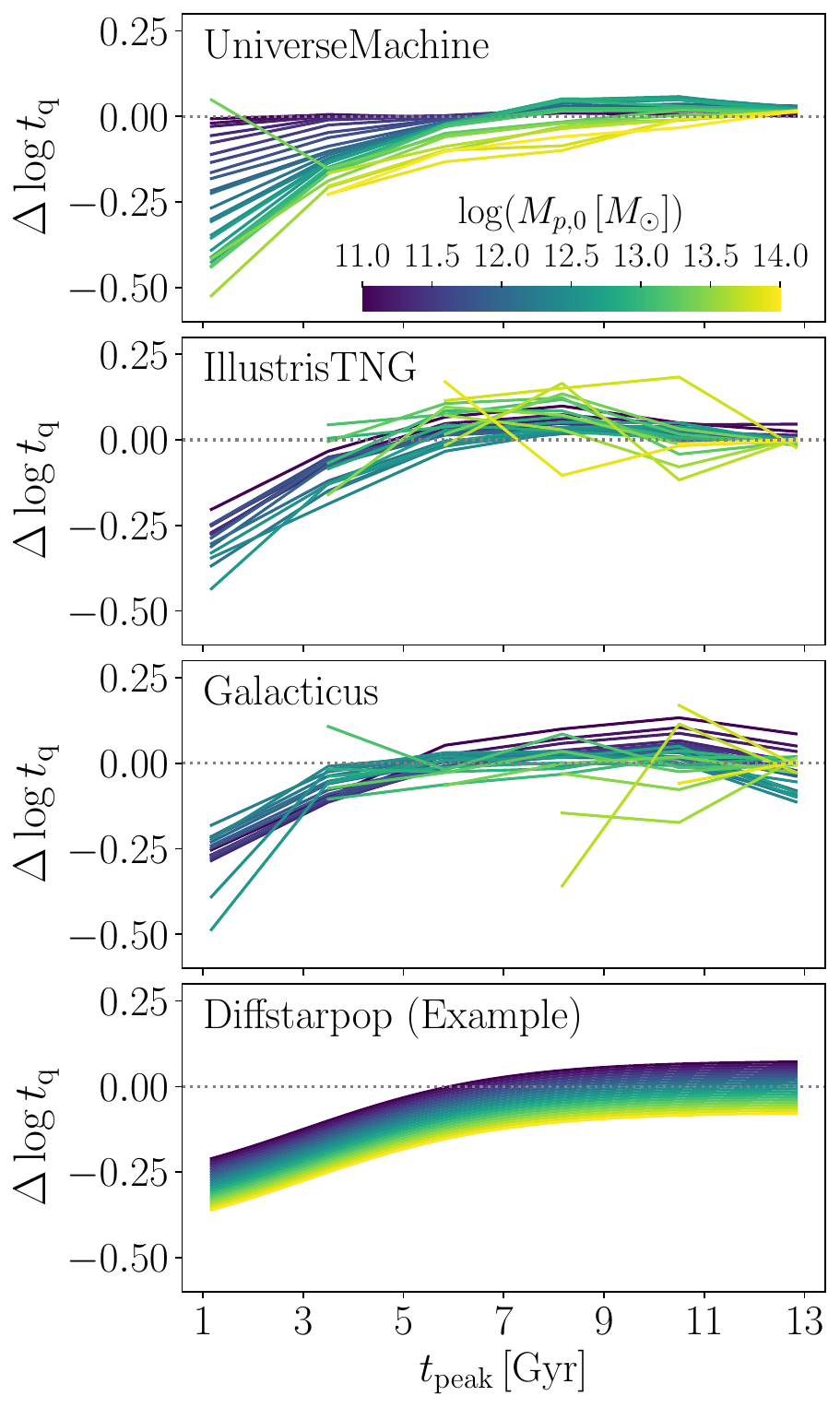}
\caption{{\bf Quenching time dependence on halo mass and $\tpeak.$} 
This figure illustrates the additional component that modifies the quenching time, $\qtime$, as a function of both the Diffmah parameter, $\tpeak$ (x-axis), and the present-day peak halo mass, $\mpeakzero$ (indicated by lines of different colors), for each halo. 
The top three panels show this dependence as measured from \um, \tng, and \galcus (all for in-situ SFH), while the bottom panel shows an example prediction from \DstarPop.}
\label{fig:uqt_tpeak}
\end{figure}

In addition, we allow early- and late-forming halos to host galaxies with different average quenching times by modifying the mean relation, $\langle \log\qtime \vert \mpeakzero \rangle$, with an additional term $\Delta \log\qtime$ that increases or decreases its value depending on both the $\tpeak$ and $\mpeakzero$ of each halo. Specifically, we use
\begin{equation}
\begin{split}
   \langle\log\qtime\vert\mpeakzero\rangle &= \sclip(\mpeakzero) + \Delta \log\qtime (\mpeakzero, \tpeak),\\[0.7em]
   \text{with:}\qquad&\Delta \log\qtime (\mpeakzero, \tpeak) = \Delta \qtimeone (\mpeakzero) + \Delta \qtimetwo (\tpeak) ,\\[0.5em]
   \text{and:}\qquad&\Delta \qtimeone (\mpeakzero) = a^{\Delta \qtime}(\mpeakzero - 12.5),\\
   &\Delta \qtimetwo (\tpeak) = S(\tpeak, x_0^{\Delta \qtime}, k^{\Delta \qtime}, y_{\rm lo}^{\Delta \qtime}, y_{\rm hi}^{\Delta \qtime}).\\
\end{split}
\end{equation}
This additional shift is controlled with a total of five free parameters, $\psi_{\rm SFH}^{\Delta \qtime} = \{a^{\Delta \qtime},x_0^{\Delta \qtime}, k^{\Delta \qtime}, y_{\rm lo}^{\Delta \qtime}, y_{\rm hi}^{\Delta \qtime}\}$. Their effect is illustrated in Figure~\ref{fig:uqt_tpeak}, where the top three
panels show this dependence as measured from \um, \tng, and \galcus, while the bottom panel shows an
example prediction from DiffstarPop. 

Overall, \DstarPop has 79 free parameters, 58 of which control the mean and standard deviation evolution with $\mpeakzero$, 16 of which control the evolution of the quenched fraction of central and satellite galaxies as a function of both $\mpeakzero$ and $\tpeak$, and 5 that control the additional shift of the mean quenching time of quenched galaxies, also as a function of $\mpeakzero$ and $\tpeak$.

\renewcommand{\thefigure}{D\arabic{figure}}
\section{Fitting DiffstarPop to Simulated Galaxy Populations}
\label{diffstarpop_fitter_appendix}

In Section~\ref{section:results} we presented fits of \DstarPop\ to the \um, \tng, and \galcus\ simulations using in-situ only star formation histories (SFHs). 
Here, we describe how these data were processed and how we performed differentiable fits to their probability density functions (PDFs). 
For complete technical details on differentiable PDF fitting, we refer the reader to previous works by \citet{hearin_etal21_diffmah},  \citet{alarcon_etal23_diffstar}, and \citet{stevanovich_etal23_diffprof}.
Below we summarize the key steps relevant to the present analysis.

\subsection{Target Data Processing}

The first step is to use \dstar\ and \dmah\ to fit each halo with $\mpeakzero > 11$ in every simulation, obtaining its best-fit SFH and mass accretion history (MAH) parameters, $\theta_{\rm SFH}$ and $\theta_{\rm MAH}$. 
From these fits, we predict the star formation history and cumulative stellar mass history for each halo using \dstar. When predicting the SFH of each galaxy with \dstar, whether for individual haloes in the target data or for model predictions with \DstarPop, we clip the \dstar\ SFHs to a minimum value of ${\rm sSFR}(t_{\rm obs}) = 10^{-12}\,{\rm yr}^{-1}$.
At a given observed redshift, we then bin halos according to their observed halo mass, $M_{\rm p}(t_{\rm obs})$, and use a histogram to compute the PDF of their stellar masses, $M_{\star}(t_{\rm obs})$, yielding
$P(M_{\star}(t_{\rm obs}) \,|\, M_{\rm p}(t_{\rm obs}))$. Similarly, we compute the specific star formation rate, ${\rm sSFR}(t_{\rm obs}) = \dot{M}_{\star}(t_{\rm obs})/M_{\star}(t_{\rm obs})$,
and construct its PDF as a function of $M_{\star}(t_{\rm obs})$, $P({\rm sSFR}(t_{\rm obs})\vert M_{\star}(t_{\rm obs}))$. 
For the sSFR distributions, we treat central and satellite galaxies separately.
We adopt 25 logarithmically spaced bins in the range $M_{\star}(t_{\rm obs}) = 10^{7} - 10^{13}\, M_{\odot}$ for the stellar-mass PDFs, and 29 logarithmically spaced bins in the range ${\rm sSFR}(t_{\rm obs}) = 10^{-13} - 10^{-8}\,{\rm yr}^{-1}$ for the sSFR PDFs. 

\subsection{Differentiable PDFs}

To enable differentiable predictions of PDFs, we further process the data so that each galaxy contributes smoothly to multiple histogram bins. 
If each galaxy contributed a discrete value of either 0 or 1 to a single bin, the gradient of any bin’s frequency with respect to an underlying quantity (e.g. $M_{\star}$) would be zero everywhere, leading to vanishing gradients with respect to the model parameters that control those quantities. This is a typical problem of discrete (or categorical) distributions, which need to be re-parametrized and made continuous (i.e. continuously relaxed) to be made differentiable \citep[e.g.][]{horowitz_etal24_diffhod}.

To address this, we assign each galaxy a probabilistic contribution across nearby bins using a \emph{triweight} kernel centered at its observed value, with a width equal to one bin width. 
The triweight function \citep[see for example Appendix E in][]{hearin_etal21_shamnet} is a compact polynomial kernel with support restricted to the interval $[\mu-3\sigma, \mu+3\sigma]$. The polynomial coefficients are tuned so that the function shares the same mean and variance as a Gaussian, and vanishes outside that range with a vanishing derivative at the endpoints, so that the triweight kernel is essentially a differentiably clipped Gaussian.
We evaluate the contribution of each galaxy to a given bin by computing the cumulative distribution function (CDF) of the triweight kernel at the bin edges and taking the difference. 
Because both the kernel and its CDF are analytic polynomials, this operation is fast and fully differentiable.

As a result, the predicted PDFs from our model differ from the true PDFs by a convolution with a triweight kernel. 
To ensure consistency, we apply the same smoothing procedure to the target (simulation) data before fitting. 
Thus, all PDFs shown throughout this paper have been computed assuming a probabilistic contribution from each individual halo, ensuring that both the model and target PDFs are differentiable and directly comparable.
    
\subsection{Additional fits to simulations with ex-situ star formation} \label{sec:exsitu}

\begin{figure*}[htbp]
  \centering
  \subfloat[\um\ stellar-to-halo-mass relation across redshift for in-plus-ex-situ SFH.\label{fig_pdf_mstar_um_inplusex}]{
    \includegraphics[width=\columnwidth]{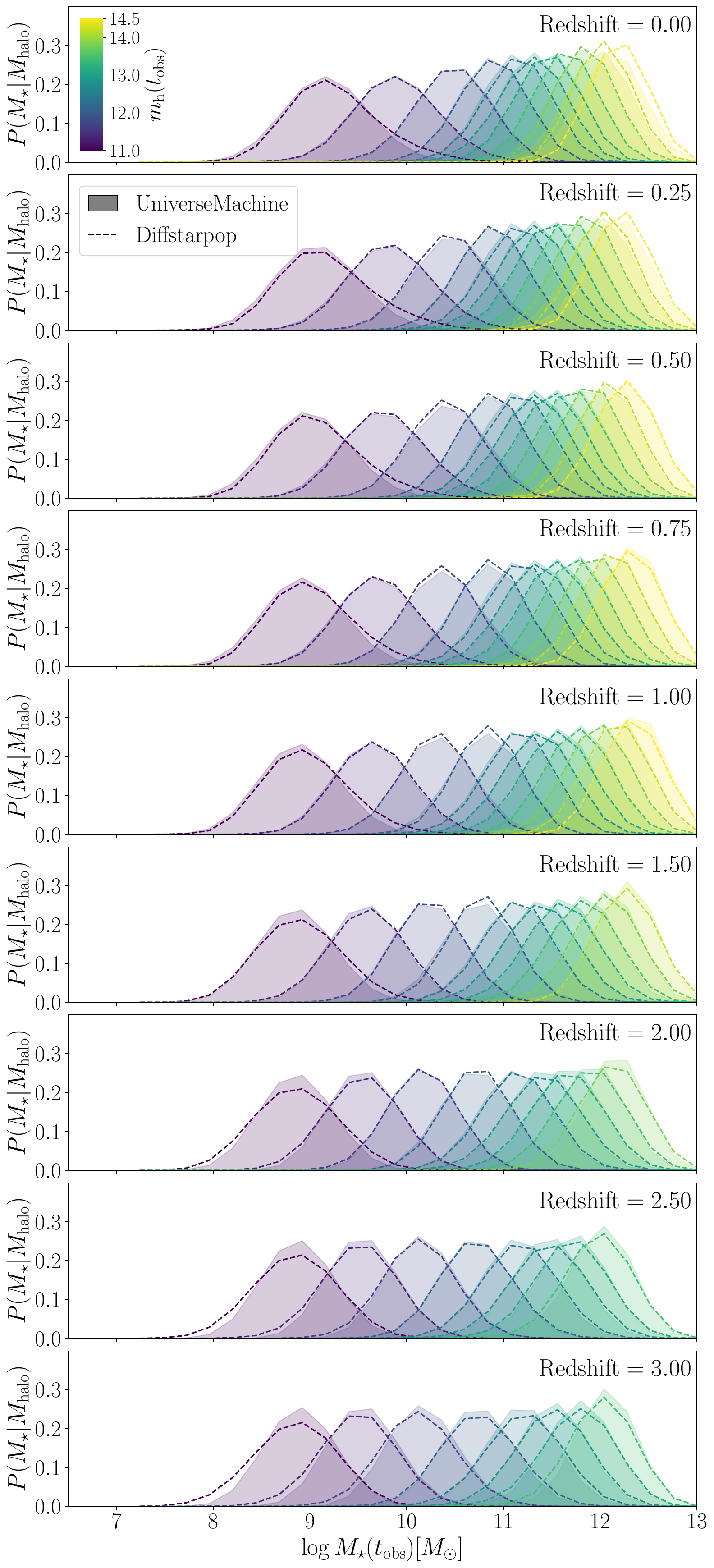}
  }\hfill
  \subfloat[\um\ sSFR PDF across redshift for centrals, for in-plus-ex-situ SFH.\label{fig_ssfr_pdf_um_inplusex}]{
    \includegraphics[width=\columnwidth]{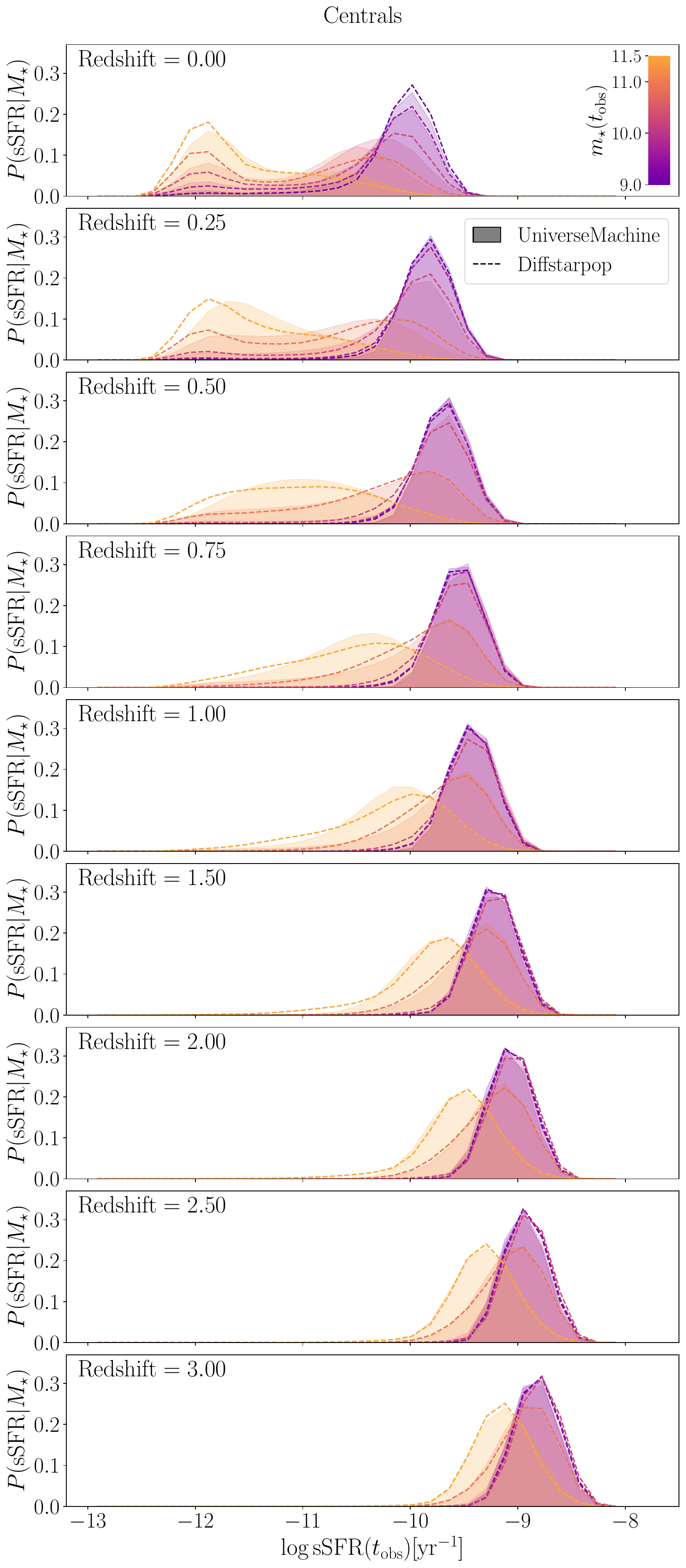}
  }\\[1ex]
  \caption{\textbf{Results for in-plus-ex-situ SFH runs of \um\ galaxies.}}
\end{figure*}

\begin{figure*}[htbp]
  \centering
  \subfloat[\galcus\ stellar-to-halo-mass relation across redshift for in-plus-ex-situ SFH.\label{fig_pdf_mstar_galcus_inplusex}]{
    \includegraphics[width=\columnwidth]{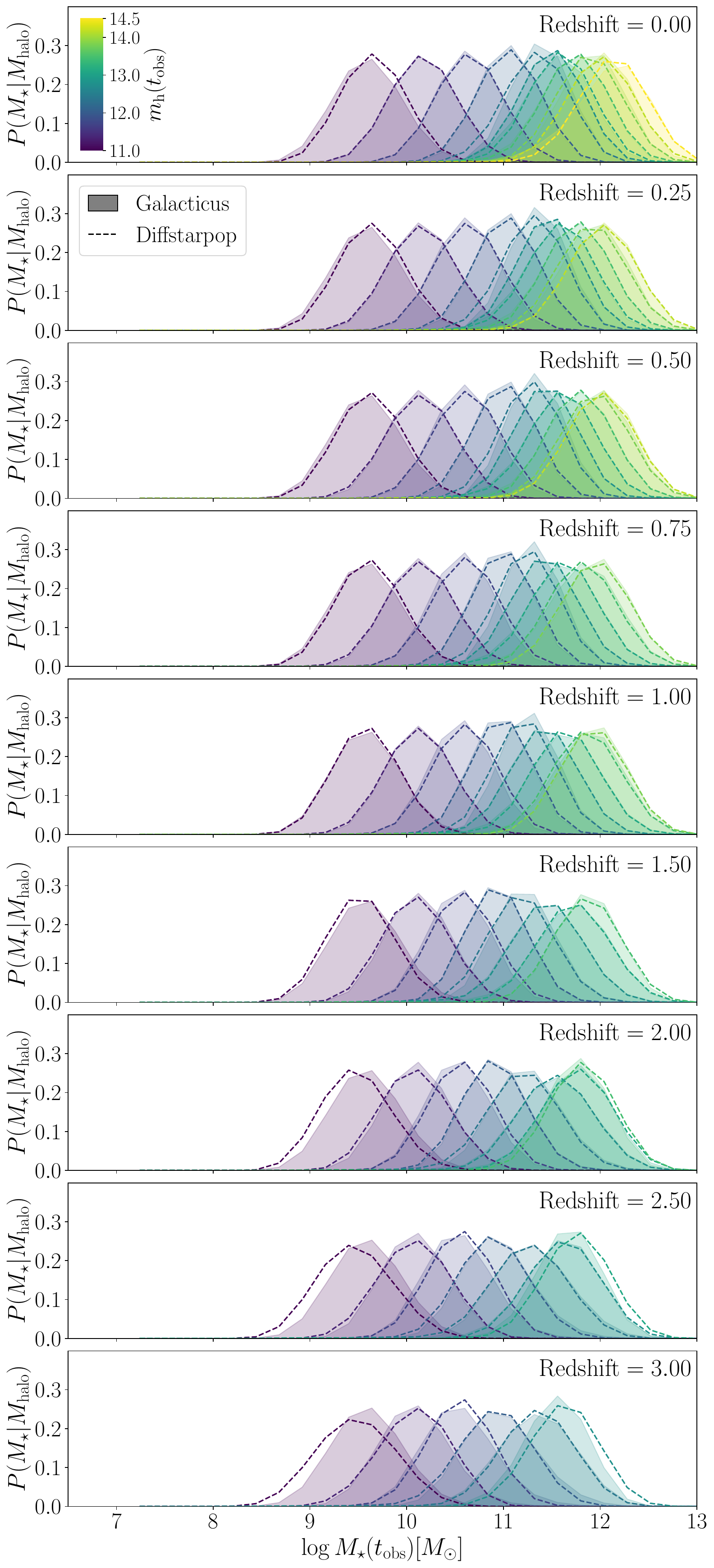}
  }\hfill
  \subfloat[\galcus\ sSFR PDF across redshift for centrals, for in-plus-ex-situ SFH.\label{fig_ssfr_pdf_galcus_inplusex}]{
    \includegraphics[width=\columnwidth]{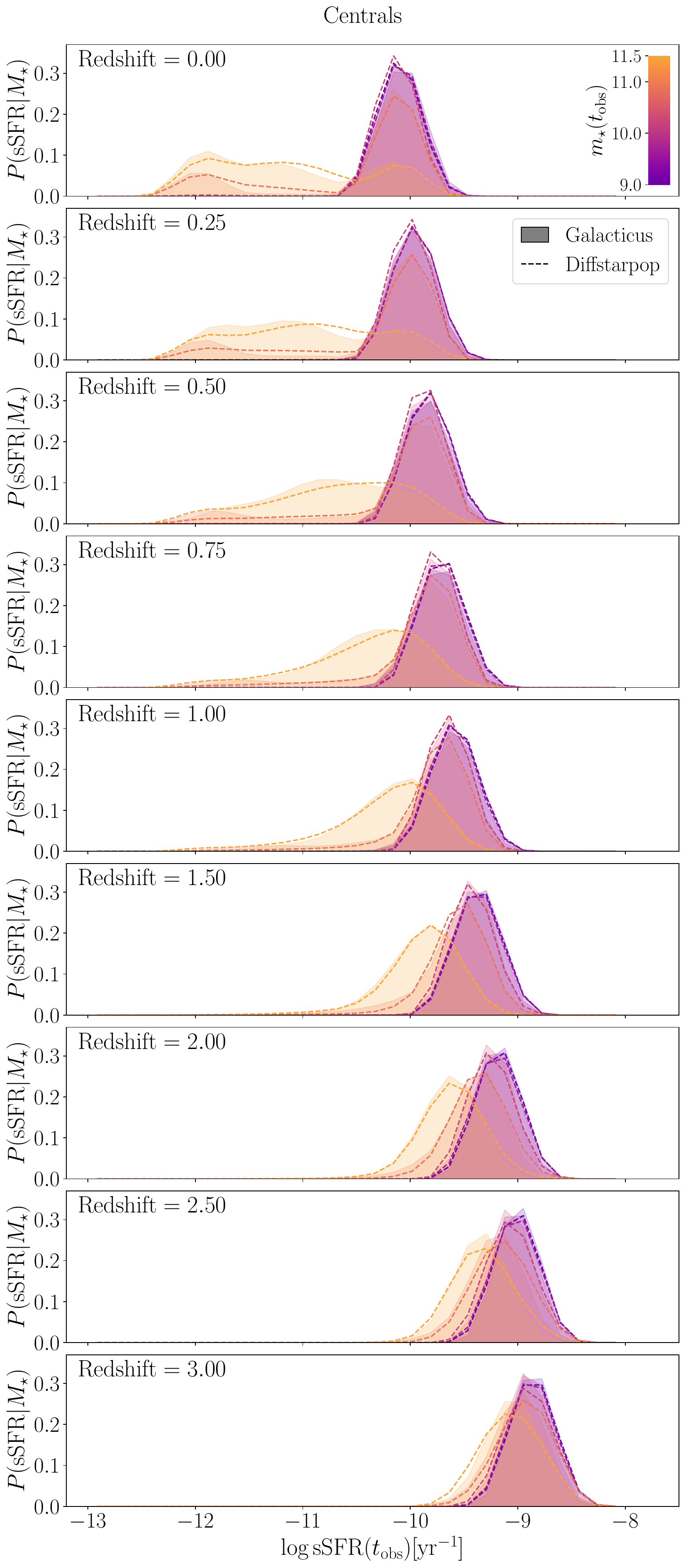}
  }
  \caption{\textbf{Results for in-plus-ex-situ SFH runs of \galcus\ galaxies.}}
\end{figure*}

Even though \dstar and \DstarPop are originally designed to model the diversity of in-situ star formation histories, here we demonstrate the flexibility of \DstarPop at describing simulations whose star formation histories include both in-situ and ex-situ components arising from mergers. Fig~\ref{fig_pdf_mstar_um_inplusex} and Fig~\ref{fig_ssfr_pdf_um_inplusex} show the results for \um in-plus-ex-situ, while Fig~\ref{fig_pdf_mstar_galcus_inplusex} and Fig~\ref{fig_ssfr_pdf_galcus_inplusex} show results for \galcus in-plus-ex-situ, showing results of stellar-to-halo-mass relation across redshift like those from Figure~\ref{fig_pdf_mstar_um}, and of sSFR PDF across redshift for
centrals as a function of stellar mass, like those from Figure~\ref{fig_ssfr_pdf_um}. Overall, we find \DstarPop is able to successfully adjust its main sequence efficiency and quenching relations to absorb the additional impact brought from ex-situ growth.

\subsection{Additional fits to sSFR of satellite galaxies} \label{sec:satellites}

In the main body and in the Appendices we have focused on results from sSFR from central galaxies. For completeness, here we show results for the PDFs of sSFR of satellite galaxies for all simulations, which have been fit simultaneously to the stellar mass and sSFR distributions from centrals.

\begin{figure*}[htbp]
  \centering
  \subfloat[Only in-situ SFH.\label{fig_ssfr_pdf_sat_um}]{
    \includegraphics[width=\columnwidth]{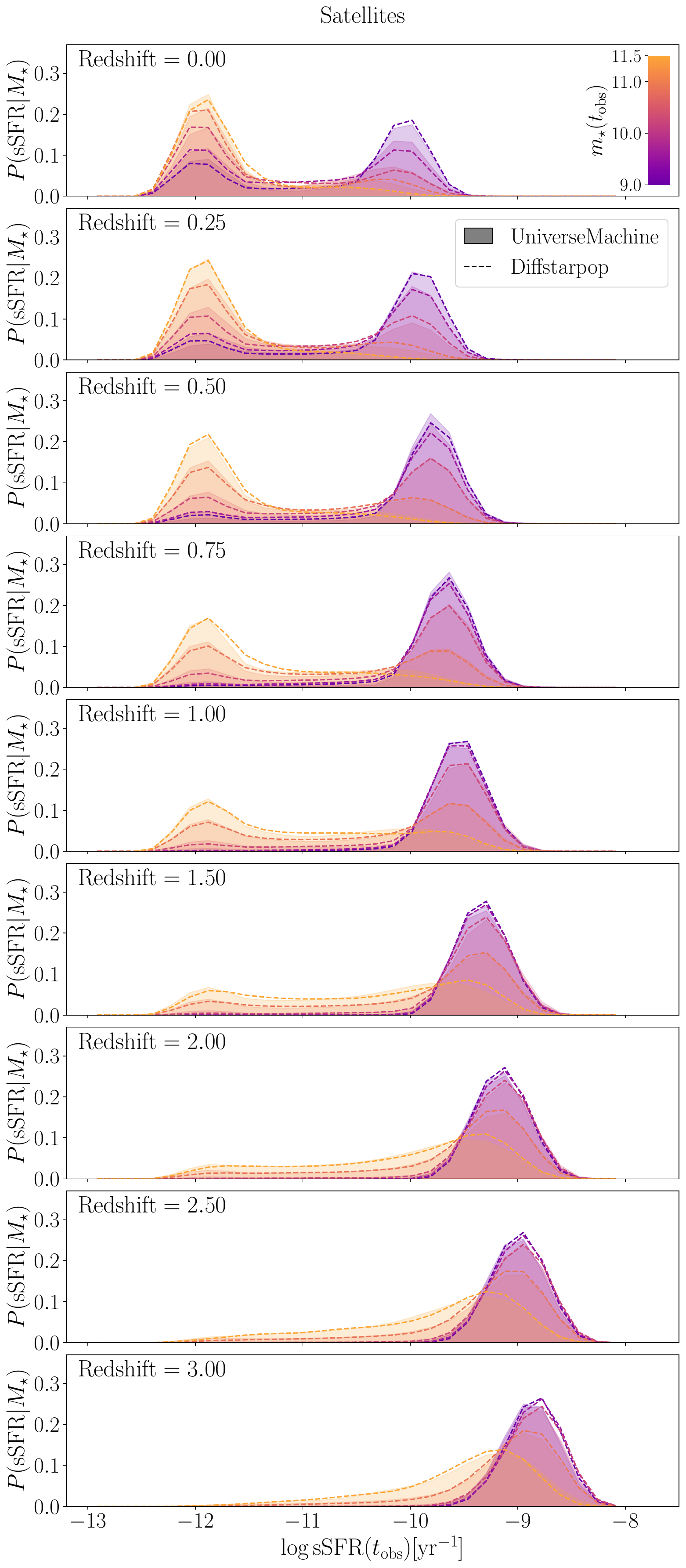}
  }\hfill
  \subfloat[Both in-situ and ex-situ contributions to SFH.\label{fig_ssfr_pdf_sat_um_inplusex}]{
    \includegraphics[width=\columnwidth]{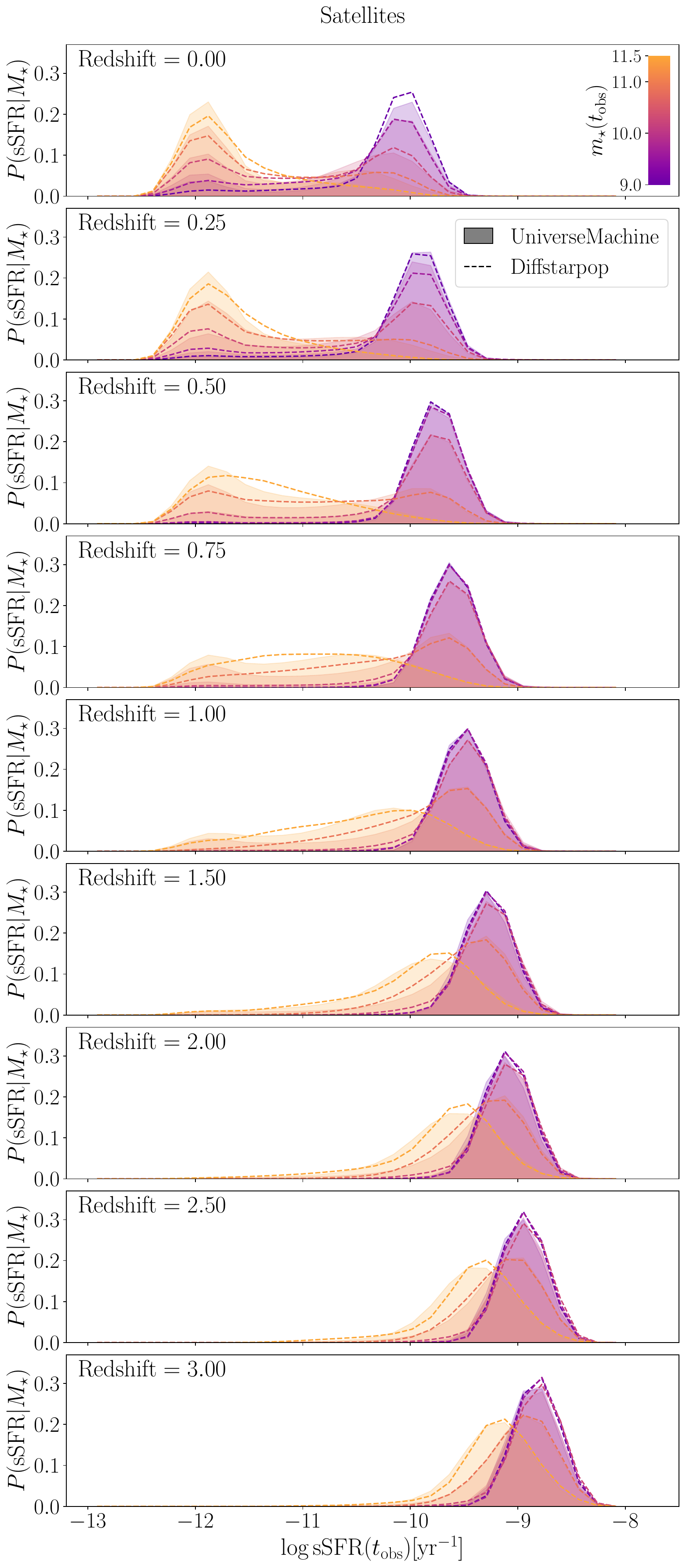}
  }\\[1ex]
  \caption{\textbf{Results of sSFR PDFs across redshift for \um\ satellite galaxies.}}
\end{figure*}

\bit
\item Fig~\ref{fig_ssfr_pdf_sat_um} shows \um for in-situ SFH.
\item Fig~\ref{fig_ssfr_pdf_sat_um_inplusex} shows \um for in-plus-ex-situ SFH.

\begin{figure*}[htbp]
  \centering
  \subfloat[Only in-situ SFH. \label{fig_ssfr_pdf_sat_galcus}]{
    \includegraphics[width=\columnwidth]{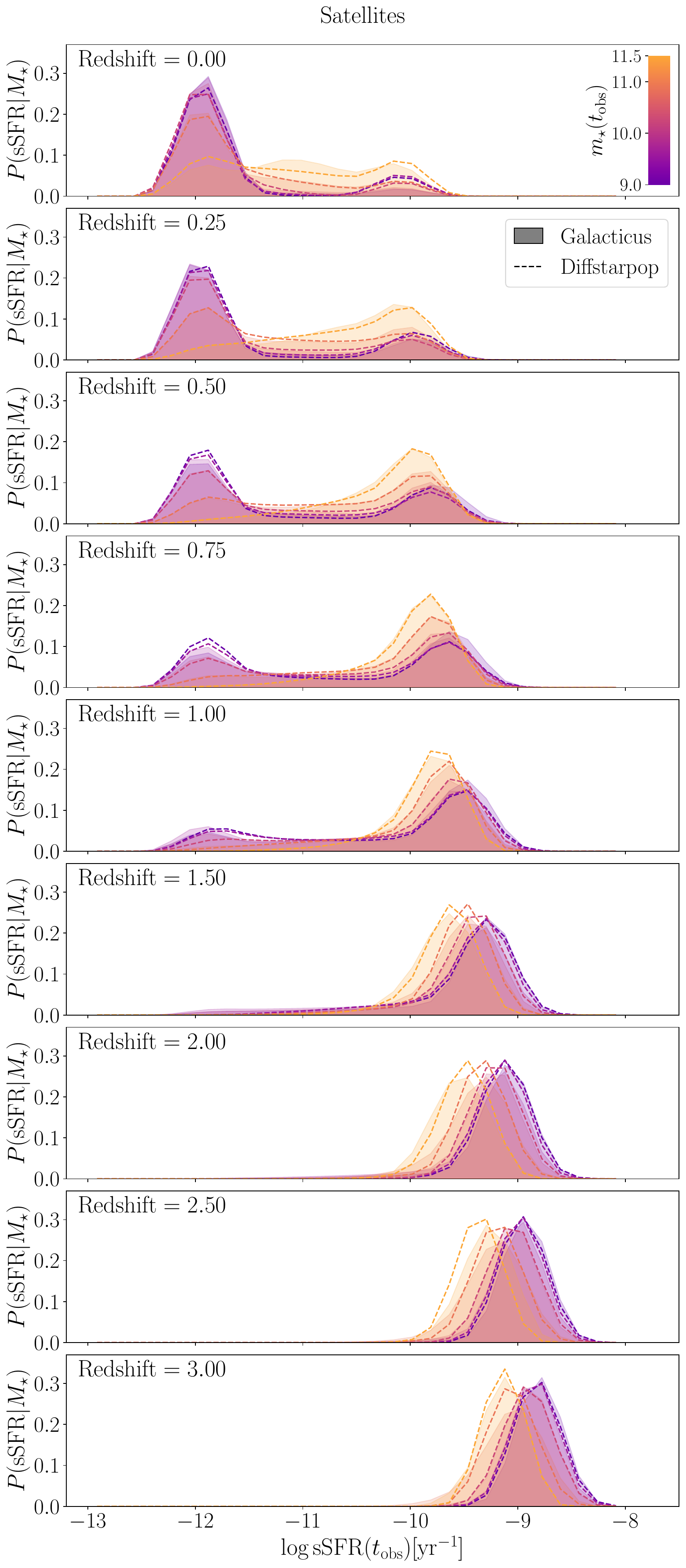}
  }\hfill
  \subfloat[Both in-situ and ex-situ contributions to SFH.\label{fig_ssfr_pdf_sat_galcus_inplusex}]{
    \includegraphics[width=\columnwidth]{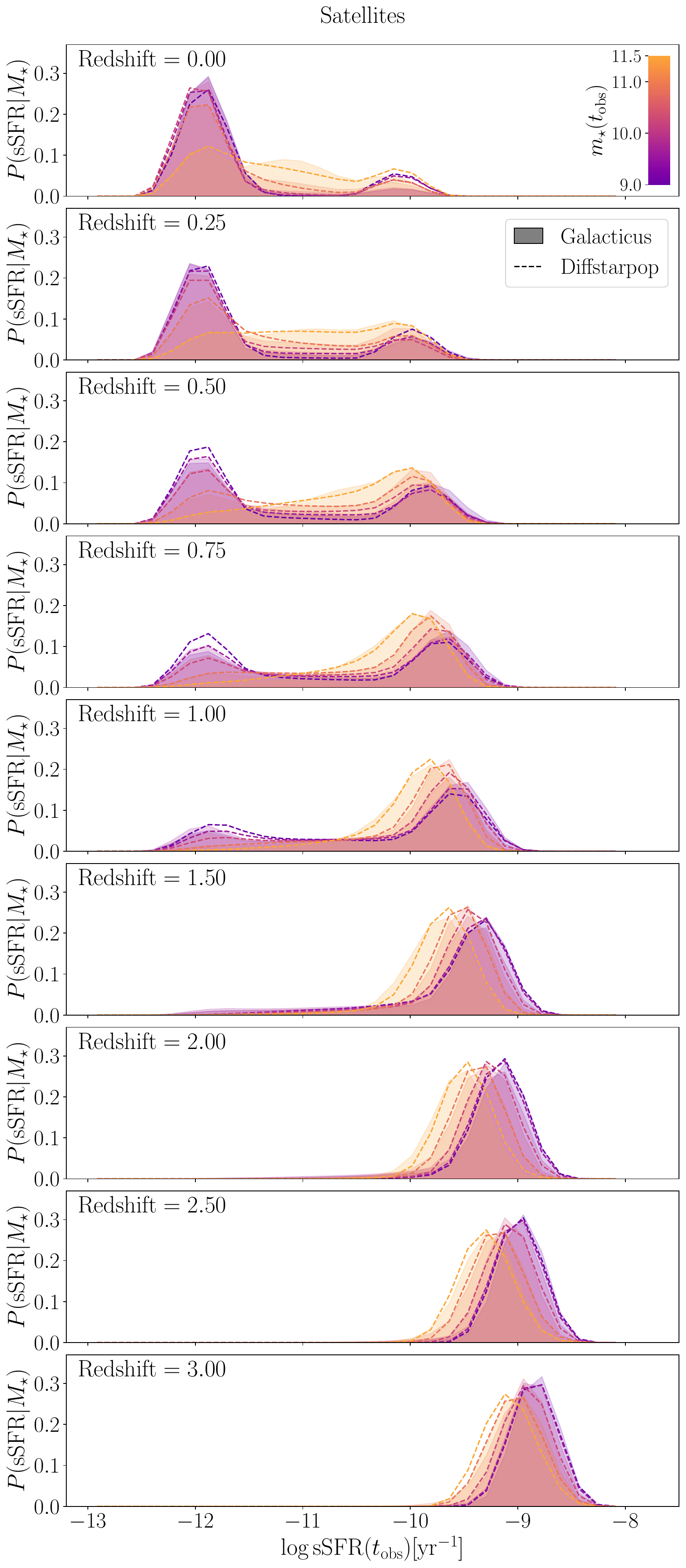}
  }

  \caption{\textbf{Results of sSFR PDFs across redshift for \galcus\ satellite galaxies.}}
\end{figure*}

\begin{figure}
\includegraphics[width=\columnwidth]{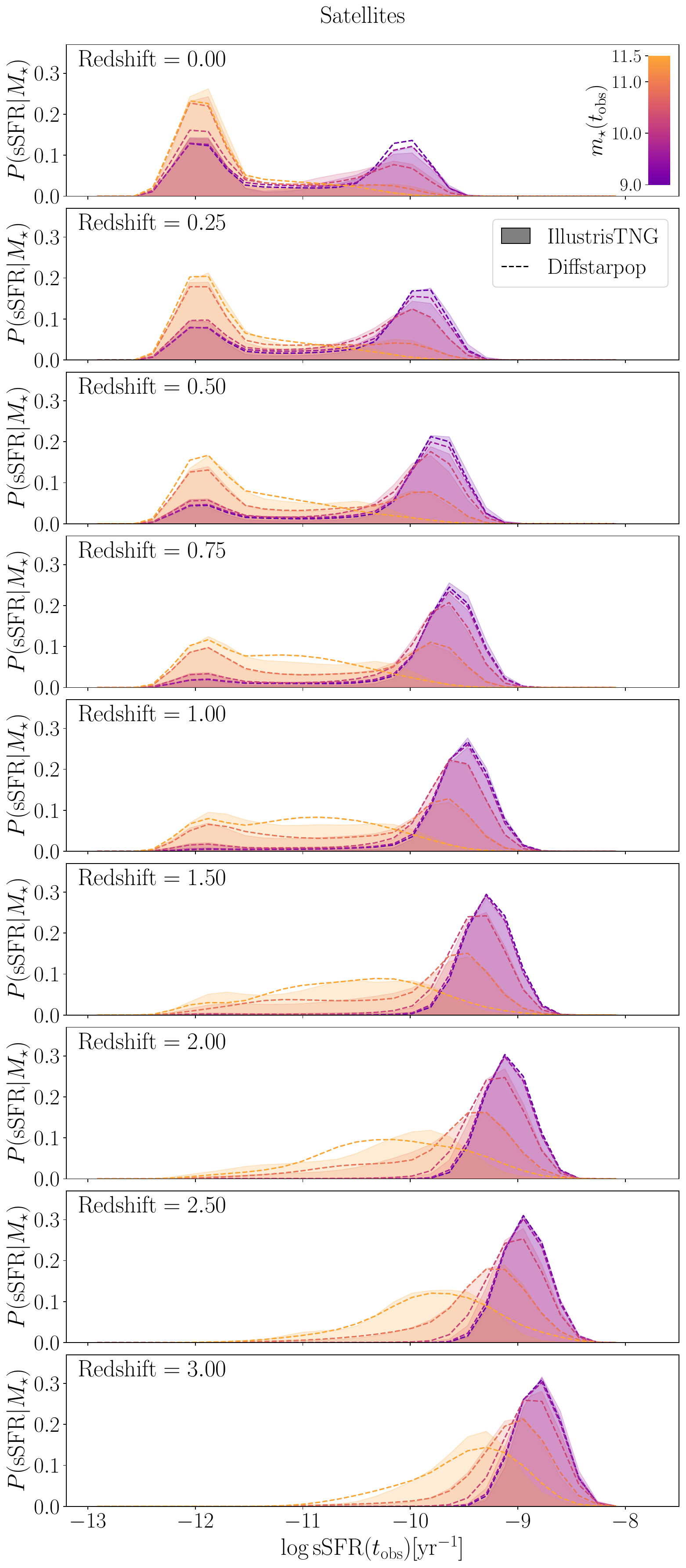}
\caption{\textbf{Results of sSFR PDFs across redshift for \tng\ satellite galaxies.}}
\label{fig_ssfr_pdf_sat_tng}
\end{figure}

\item Fig~\ref{fig_ssfr_pdf_sat_galcus} shows \galcus for in-situ SFH.
\item Fig~\ref{fig_ssfr_pdf_sat_galcus_inplusex} shows \galcus for in-plus-ex-situ SFH.
\item Fig~\ref{fig_ssfr_pdf_sat_tng} shows \tng for in-situ SFH.
\eit

\subsection{Best-fit predictions of \dstar parameters from \DstarPop} \label{sec:bestfit_diffpop}

\begin{figure*}
\centering
\includegraphics[width=0.92\linewidth]{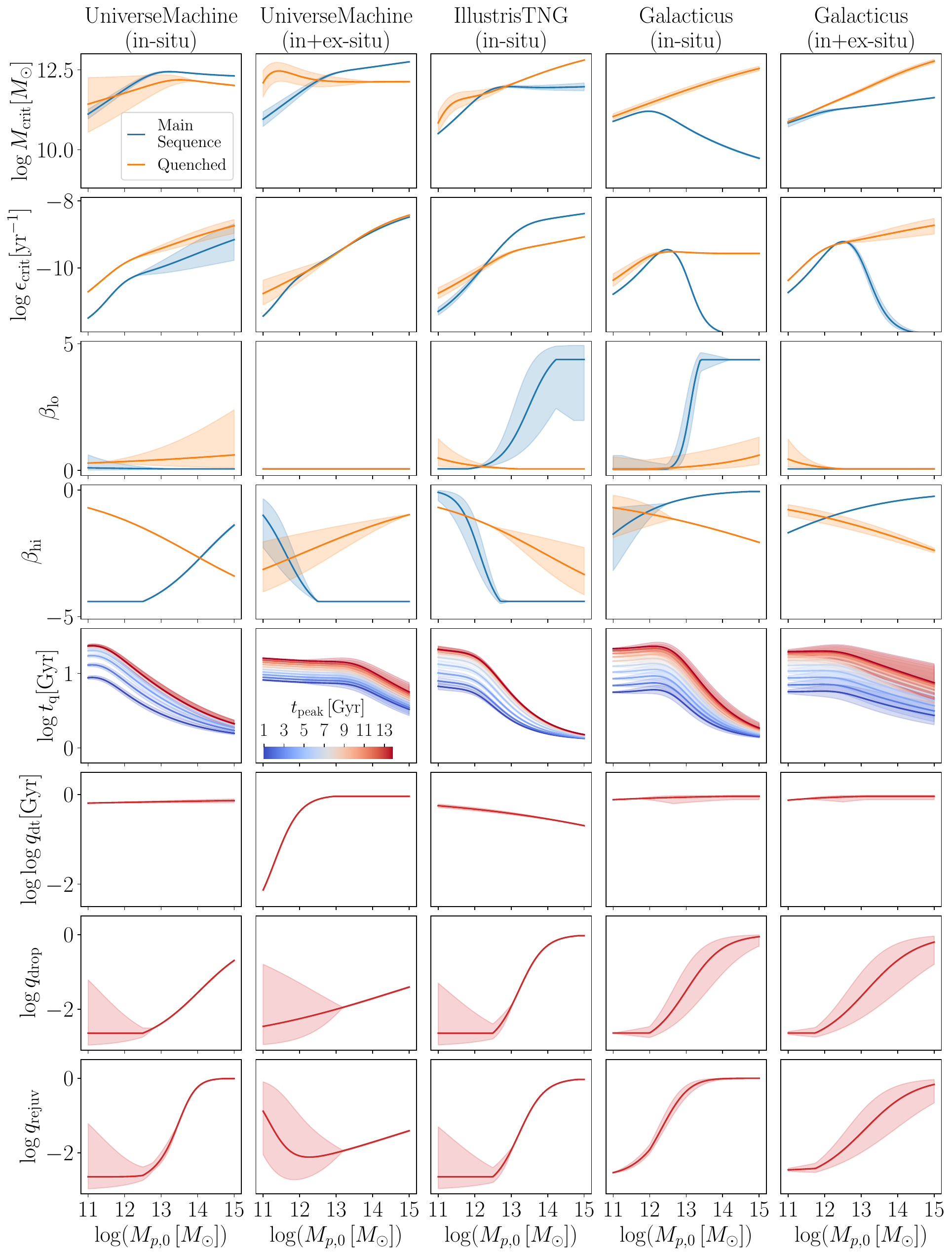}
\caption{{\bf \DstarPop best-fit relations of \dstar parameters.} The rows show the mean relations for each of the 8 \dstar parameters as a function of the present-day halo mass $\mpeakzero$, for each of the 5 simulations studied in this work (shown in different columns). The lines show the mean relations, and the shaded regions show the spread in the parameters (which is Gaussian in unbounded space, but can look skewed here). The top four rows show the main sequence parameters, both for the main sequence galaxies (blue lines) and for the quenched galaxies (orange lines). The bottom four rows show the quenched parameters, which are only relevant for the quenched population. For the logarithmic quenching time row, $\log \qtime$, we additionally show the dependence on the Diffmah parameter $\tpeak$ (lines with different colors).}
\label{fig_diffpop_bestift_relations}
\end{figure*}

\begin{figure}
\includegraphics[width=\columnwidth]{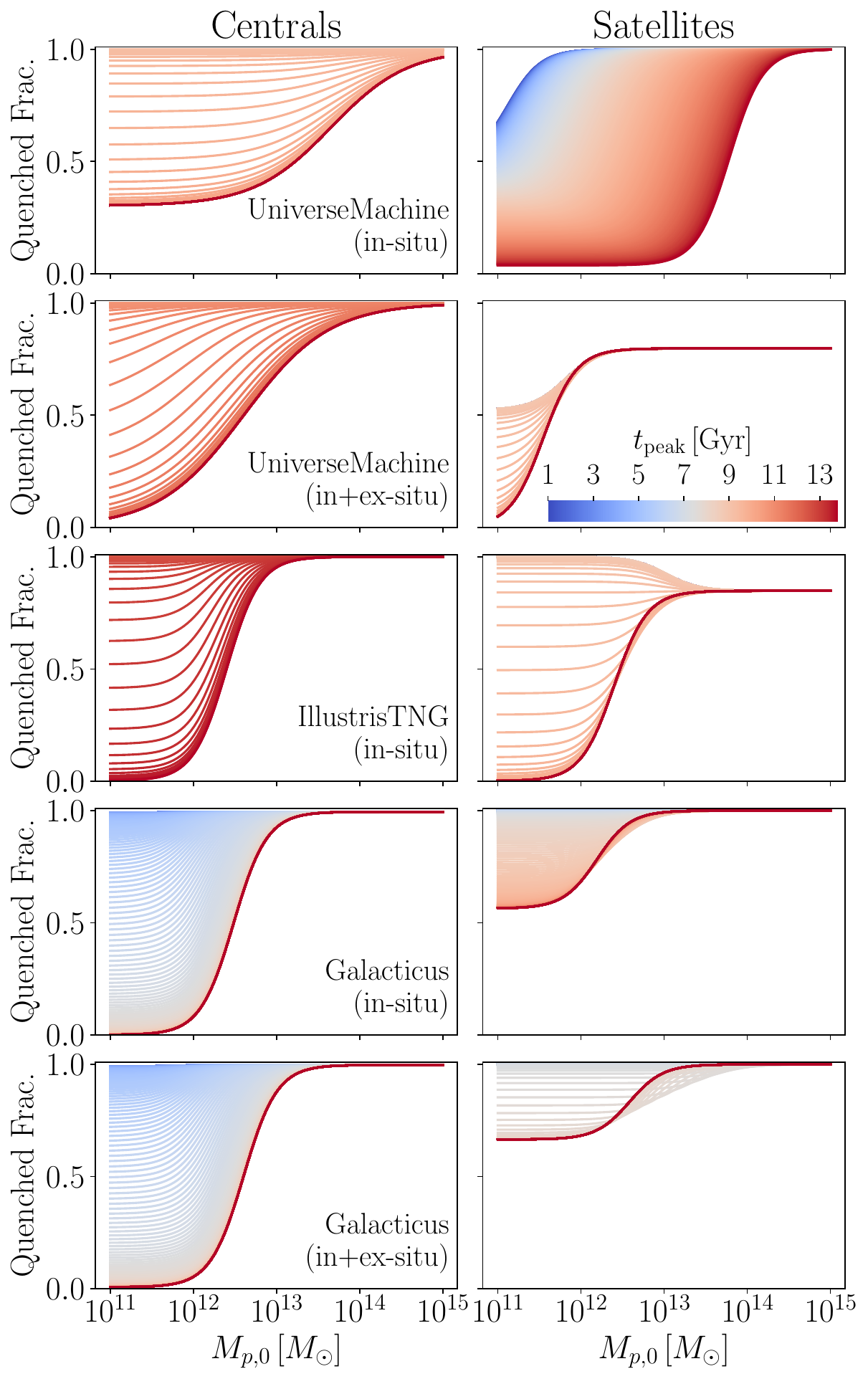}
\caption{{\bf \DstarPop best-fit quenched fractions.} The best-fit \DstarPop characterizing galaxy quenching for each of the five simulations analyzed. The y-axis represents the fraction of objects that experience a quenching event at some time $\qtime<t_0.$ The fits are obtained by modeling the probability distributions of stellar mass and specific star formation rate described in \S\ref{section:results}.}
\label{fig_diffpop_bestift_fquench}
\end{figure}

Here we present the predicted best-fit \DstarPop\ relations governing the main-sequence and quenched Gaussian components for each of the \dstar\ parameters as a function of present-day halo mass $\mpeakzero$ (Figure~\ref{fig_diffpop_bestift_relations}), as well as for the quenched fractions that mix them (Figure~\ref{fig_diffpop_bestift_fquench}). These relations are obtained by fitting \DstarPop\ to the target stellar-mass and sSFR distributions of each simulation.

\bigskip

\end{document}